\documentclass[a4paper,12pt]{article}

\usepackage[english]{babel} 
\usepackage[utf8x]{inputenc} 
\usepackage[T1]{fontenc}

\usepackage[a4paper,top=2cm,bottom=2cm,left=3cm,right=3cm,marginparwidth=1.75cm]{geometry} 
\usepackage[font=smaller]{caption}

\usepackage{xcolor}
\usepackage{amsmath,physics,amssymb} 
\usepackage{graphicx} 
\usepackage[colorlinks=true, allcolors=blue]{hyperref} 
\usepackage{caption} 
\usepackage{subcaption} 
\usepackage[noabbrev]{cleveref} 
\usepackage[title]{appendix} 
\usepackage{booktabs} 
\usepackage{siunitx} 
\usepackage{multirow} 
\usepackage[numbers,sort&compress]{natbib} 
\usepackage[shortlabels]{enumitem}
\usepackage[normalem]{ulem}

\renewcommand{\vec}[1]{\boldsymbol{#1}}
\title{Cooperative success in epithelial public goods games} 
\author{Jessie Renton\thanks{Corresponding author: jessica.renton.16@ucl.ac.uk} \qquad Karen M.\ Page \\ 
\emph{Department of Mathematics, University College London,}\\\emph{Gower Street, London WC1E 6BT, UK}}
\date{}

\begin{document} 
\maketitle
\begin{abstract} Cancer cells obtain mutations which rely on the production of diffusible growth factors to confer a fitness benefit. These mutations can be considered cooperative, and studied as public goods games within the framework of evolutionary game theory. The population structure, benefit function and update rule all influence the evolutionary success of cooperators. We model the evolution of cooperation in epithelial cells using the Voronoi tessellation model. Unlike traditional evolutionary graph theory, this allows us to implement global updating, for which birth and death events are spatially decoupled. We compare, for a sigmoid benefit function, the conditions for cooperation to be favoured and/or beneficial for well-mixed and structured populations. We find that when population structure is combined with global updating, cooperation is more successful than if there were local updating or the population were well-mixed. Interestingly, the qualitative behaviour for the well-mixed population and the Voronoi tessellation model is remarkably similar, but the latter case requires significantly lower incentives to ensure cooperation.
    
\end{abstract}

\begin{small} \textbf{\emph{Keywords--- }} multiplayer games, cooperation, evolutionary game theory, Voronoi tessellation, epithelial automata
\end{small}

\section{Introduction}
\subsection{Cooperation between cancer cells}
Oncogenesis is a process of somatic evolution. In order to become cancerous there are certain key mutations which cells must obtain, corresponding to the hallmarks of cancer \cite{Hanahan2000,Hanahan2011}. Evolutionary game theory provides a framework for modelling mutations which have a fitness effect beyond the cell itself. For example, certain mutations can be considered cooperative, in that they invoke a cost to the cell which is recuperated as a shared benefit. This is evident when the benefit relies on the production of a diffusible growth factor \cite{Jouanneau1994,Axelrod2006}, as is the case for a number of the hallmarks of cancer, such as self-sufficiency in growth signalling and sustained angiogenesis. The Warburg effect, whereby tumour cells metabolise through glycolysis even when oxygen is abundant \cite{Warburg1956}, can also be considered cooperative \cite{Archetti2014}.

Cooperative mutations benefit the population as a whole; however, it is often the case that defection (e.g.\ not producing growth factor) results in higher individual fitness. This is because the defector shares in the benefits without paying any fitness costs associated with cooperating. Understanding the conditions under which cooperation can evolve, despite the incentive to defect, has been a topic of extensive study within evolutionary game theory \mbox{\cite{Nowak2006,Ohtsuki2006a,Allen2016}}.

Cooperation is usually considered to be a desirable outcome. For example, within the social sphere or amongst healthy constituent cells of a multicellular organism. Cooperation between cancer cells, however, can drive tumour growth \mbox{\cite{Marusyk2014}}. This is of course detrimental to the patient, and thus, disrupting cooperation between cancer subclones, possibly by exploiting its evolutionary weaknesses, could be an important avenue for treatment \mbox{\cite{Archetti2013,Zhou2017}}.

\subsection{Public goods games}
Applications of evolutionary game theory to model cancer evolution have mainly focussed on two-player games, whereby cells participate in multiple pairwise interactions within the population \cite{Tomlinson1997b,Basanta,Hummert2014}. Interactions between cancer cells however, tend to happen in groups. For example, a cell producing a growth factor will provide a benefit to other cells within its diffusion range. These types of mutations are thus better represented as multiplayer public goods games (PGGs) \cite{Archetti2012}, played between producer (cooperator) and non-producer (defector) cells. The former produce growth factor at a fixed cost to their fitness. Both producers and non-producers receive a fitness benefit as a function of the frequency of producers in their interaction neighbourhood.  

The most common PGG, known as the N-person prisoner's dilemma (NPD), uses a linear benefit function \cite{Hauert2002,Santos2008}. However, non-linear benefit functions may be more realistic \cite{Archetti2015,Archetti2020}, and can lead to much richer dynamics, even for well-mixed populations. An example is the volunteer's dilemma (VD), which defines the benefit as a Heaviside step function \cite{Bach2001,Bach2006,Archetti2009a,Archetti2009}. 

A sigmoid benefit function has been proposed as an appropriate model for growth factor production. Experiments on neuroendocrine pancreatic cancer cells \emph{in vitro} have found sigmoid dependence of proliferation rates on the concentration of growth factor IGF-II \cite{Archetti2015}. Furthermore, such a function is relatively general, with both the NPD and VD arising as extreme cases \cite{Archetti2011}. 

\subsection{Population structure and update rules}
Most cancers originate in epithelia. These are tissues formed of sheets of cells, which are approximately polygonal on their apical surfaces. It is important to take into account this population structure when modelling the evolutionary dynamics. For both two-player cooperation games \cite{Ohtsuki2006,Nowak2010} and multiplayer PGGs \cite{Pena2016}, cooperators tend to have greater success in structured populations, as compared to well-mixed ones, because they are able to form mutually beneficial clusters. 

Evolution on structured populations is usually modelled within the framework of evolutionary graph theory \mbox{\cite{Lieberman2005}}, in which the population is represented as a fixed graph. Epithelial cells tend to have six neighbours on average, and thus can be represented as a hexagonal lattice. Introducing more realistic population structures, with small variation in neighbour number, does not have a significant impact on evolutionary outcomes \cite{Archetti2016,Renton2019}. 

The success of cooperation is also dependent on the update dynamics. Within evolutionary graph theory, the population evolves according to an update rule. In general, update rules can be divided into two categories: local and global \cite{Nathanson2009}. 

\subsubsection{Local updating}
A local update involves a spatial relationship between birth and death events. Evolutionary graph theory usually requires a local update rule in order to maintain the fixed graph structure. Two commonly used local update rules are defined as follows:
\begin{itemize}
    \item \emph{birth-death:} a cell is selected to divide with probability proportional to fitness; one of its neighbours is chosen to die uniformly at random.
    \item \emph{death-birth:} a cell is chosen to die uniformly at random; one of its neighbours is selected to divide with probability proportional to fitness
\end{itemize}
In both cases the offspring of the dividing cell occupies the empty site left by the dead cell \cite{Zukewich2013}. The choice between these update rules has a substantive effect on evolutionary outcomes. For example, consider a two-player prisoner's dilemma game and a population represented by a regular graph. Cooperation can be favoured for a death-birth update rule, so long as the benefit is high enough. For the birth-death update, however, as is the case with a well-mixed population, cooperation is only favoured for an infinitely large benefit \cite{Ohtsuki2006}. 

These update rules are sometimes referred to as BD-B (birth-death with selection on birth) and DB-B (death-birth with selection on birth) to emphasise that selection is acting on birth. Alternative update rules, for which selection acts on death, can then be referred to as BD-D and DB-D \cite{Masuda2009}. In this paper, we limit ourselves to the case where selection acts on birth, thus we do not use this notation to differentiate the two cases.

\subsubsection{Global updating}
Under a global update rule there is no spatial dependence between birth and death events, thus cells are selected to reproduce and die from the population as a whole. Global updating is generally seen for well-mixed populations, or when populations are organised in phenotype space \cite{Antal2009} or by sets \cite{Tarnita2009a}. 

Within evolutionary graph theory the \emph{shift} update rule is an example of global updating. In this case a cell is chosen to divide with probability proportional to fitness, and a second cell is chosen to die uniformly at random. A path is then selected on the graph which connects the two. Cells are shifted along this path until there is an empty node next to the dividing cell for its progeny to occupy. This kind of update works well on a one-dimensional lattice \cite{Allen2012}, and promotes cooperation, even compared to the death-birth update. However, it becomes more complex in two-dimensions \cite{Pavlogiannis2015}, because division causes cellular rearrangement at a distance from the event.

\subsubsection{Epithelial structure and dynamics}
Evolutionary graph theory has several shortcomings for modelling invasion processes in epithelia. Firstly, it assumes that the population can be represented by a static graph, whereas epithelia are dynamic structures. Secondly, as we have discussed, 
introducing global update rules into evolutionary graph theory presents challenges to the modelling framework \mbox{\cite{Pavlogiannis2015}}.

The question then arises as to which update rule is most realistic for an epithelium. This will depend on the extent to which death and division processes are spatially coupled. For homeostatic tissues it is likely that contact inhibition, the phenomenon whereby cells stop proliferating at high density, plays an important role in maintaining the population size \mbox{\cite{Mesa2018}}. The death-birth update rule could be an appropriate model when contact inhibition is very strong, as tissue density is likely to be low near a recent death. Conversely, a global update rule is likely to be more realistic when contact inhibition is weaker and thus there is less spatial dependence between death and division.

The death-birth and decoupled update rules represent extreme cases of spatial coupling between division and death. In this paper we focus on global updating, as the death-birth update rule, along with other local update rules, has been extensively studied within evolutionary graph theory \mbox{\cite{Ohtsuki2006,Maciejewski2014a,Pena2016,Allen2016}}. In future work, we will consider the spectrum of spatial coupling that can arise in a tissue due to contact inhibition, and how this affects the evolution of cooperation.

In line with our previous work \mbox{\cite{Renton2019}}, we use the Voronoi tessellation (VT) model \mbox{\cite{Meineke2001,VanLeeuwen2009}} to represent epithelial dynamics. Unlike traditional evolutionary graph theory models, the tissue structure is dynamic and cells are able to divide and die independently. It is thus straightforward to spatially decouple birth and death, and we are able to introduce a global form of updating, we call the decoupled update rule. In \mbox{\cite{Renton2019}}, we used this framework to analyse the two-player prisoner's dilemma game, finding that cooperation was more successful for the decoupled update rule, than for a death-birth update rule. The present paper extends these results to a wide range of multiplayer public goods games, as well as deriving general results for global update rules.

We aim to extend the range of applicability of quasi-analytical methods from evolutionary game theory to more realistic tissue models. We have chosen to use the VT model, because it uses a very simple force law and, as a cell-centre model, naturally provides the graph structure needed for evolutionary games \mbox{\cite{Meineke2001}}. Furthermore, unlike cellular automata models, cell division leads only to local topological changes. The VT model has been used to represent cellular dynamics in colonic and intestinal crypts, including for models of invasion \mbox{\cite{Mirams2012,Romijn}}. Other tissue models, such as the vertex model \mbox{\cite{Farhadifar2007}}, could also be appropriate for our purposes.

The version of the VT model we use represents a simple epithelium\footnote{A simple epithelium is formed of a single layer of cells, whereas a stratified epithelium is multilayered. } as a two-dimensional structure. Thus our results are mostly relevant to the early stages of tumorigenesis or field cancerization \mbox{\cite{Curtius2017}} in simple epithelia. While models of later stage tumour evolution would be more appropriately modelled in three dimensions, two-dimensional models, such as the VT model, can still be useful in the first instance.

\subsection{Measures of mutant success}
\label{sec:conditions_for_success}
For stochastic evolutionary games without mutation, we can compare the success of different strategies by calculating fixation probabilities. Here we consider the dynamics of two cell types: $A$ and $B$. The fixation probability $\rho_X$ is then defined as the probability that a single initial mutant $X$ will eventually take over the entire population. We consider two measures for the success of an $A$ mutant \cite{Zukewich2013,Maciejewski2014a}:
\begin{itemize}
    \item $A$ is a \emph{beneficial} mutation when $\rho_A>\rho_0$. Here $\rho_0=1/Z$ is the fixation probability for a neutral mutant and $Z$ is the population size.
    \item $A$ is \emph{favoured} by selection, or has an evolutionary advantage, when $\rho_A>\rho_B$. This is equivalent to the condition that the equilibrium frequency of $A$ is greater than a half when mutation is allowed ($A$ is the dominant strategy).
\end{itemize}

In general, these conditions are not equivalent, thus it is possible for a mutation to be beneficial but not favoured, or vice versa. One or the other condition might be more relevant to quantifying mutant success depending on the circumstances. Furthermore, under certain circumstances these two conditions are equivalent \cite{Maciejewski2014a}. 

The remainder of this paper explores conditions under which a mutation is beneficial and/or favoured. We begin in \Cref{sec:PGG} by setting out the mathematical formalism for multiplayer evolutionary games, focussing particularly on PGGs played between cooperators and defectors. \Cref{sec:favoured_coop} then introduces the $\sigma$-rule, which is used to determine whether a strategy is favoured. We outline several known results on graphs with local update rules, as well as deriving results for a birth-death and shift update rule on a cycle. We then derive the conditions for favourability on a general population structure with global updating. In \Cref{sec:beneficial_coop} we derive a similar rule, but for a strategy to be beneficial. In \Cref{sec:epithelium} we apply this theory to consider conditions for cooperator success in an epithelium, using spatial statistics calculated through simulation of the Voronoi tessellation model. Finally, in \Cref{sec:discussion}, we discuss the implications of our work for the evolution of cooperative public goods in epithelia and make some remarks on the different significance of beneficial and favourable mutants.

\section{Evolutionary dynamics of multiplayer games}
\label{sec:PGG}
We consider an arbitrary multiplayer game with two strategies, $A$ and $B$. Players interact in groups of size $N=k+1$, and obtain payoffs $a_{j,k}$ and $b_{j,k}$ respectively, where $j$ is the number of $A$ co-players and $k$ is the total number of co-players. For a graph-structured population, the co-players are direct neighbours. The fitness of each individual is then defined as $1+\delta a_{j,k}$ or $1+\delta b_{j,k}$, where $\delta$ is the selection strength parameter. 

The population evolves according to a Moran process \cite{Moran1958}, i.e.\ at each time-step one individual dies and another reproduces, thus keeping the population size, $Z$, constant. How these individuals are chosen is determined by the update rule. We consider cases where reproduction, but not death, is dependent on fitness.

Many of the results we derive in the following sections are for general games, however we are focussed on PGGs played between producer/cooperator cells ($C$) and non-producer/defector cells ($D$). These games are defined by a benefit function $b\cdot\beta(x)$ and a cost function which we take to be constant $c$, with $b>c$. Here $x$ is the proportion of cooperators in a cell's interaction group. Thus cooperator and defector payoffs are defined respectively as
\begin{align}
    a_{j,k} = b\cdot\beta\left(\frac{j+1}{k+1}\right) - c\, , && b_{j,k} = b\cdot\beta\left(\frac{j}{k+1}\right)\, . \label{eq:pgg}
\end{align}
In order to ensure that the payoff is higher when all players cooperate than when no players cooperate we enforce the condition $b\cdot\beta(N) - c \ge b\cdot\beta(0)$. Often this is done by setting $c=1$, $b>1$, $\beta(N)=1$ and $\beta(0)=0$.

\begin{figure}[h!]
    \centering
    \includegraphics{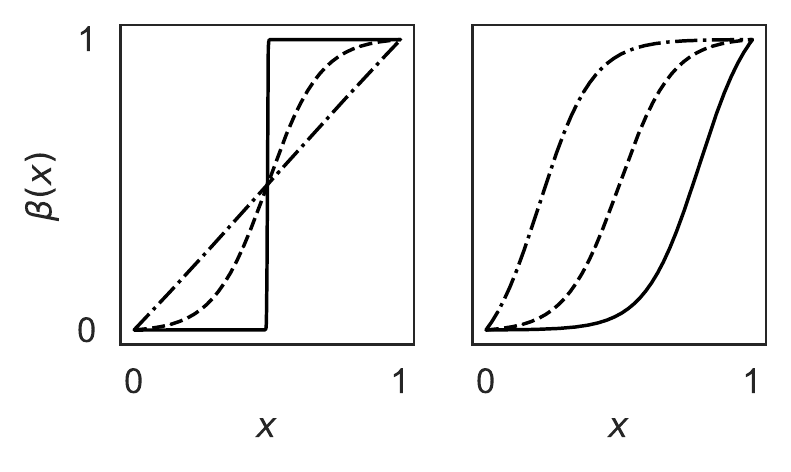}
    \caption{Logistic benefit function. \emph{Left panel:} $h=0.5$; $s=1$ (dash-dot), $s=10$ (dash) and $s=1000$ (solid). We can regain the limiting cases by letting $s\to 0$ (NPD) or $s\to\infty$ (VD). \emph{Right panel:} $s=10$; $h=0.2$ (dash-dot), $h=0.5$ (dash), $h=0.8$ (solid). }
    \label{fig:logistic_benefit}
\end{figure}

The NPD and VD can both be defined in this form by specifying the benefit functions:
\begin{align}
    \beta(x) &= x &\text{(NPD)} \label{eq:NPD}\\
    \beta(x) &= \Theta(x-\tilde x) \,, &\text{(VD)} \label{eq:VD}
\end{align}
where $\Theta(x)$ is the Heaviside step function and $\tilde x$ is the minimum proportion of cooperators required to obtain the benefit. Furthermore we can define a general sigmoid benefit function: 
\begin{equation}\label{eq:sigmoid_benefit} 
    \beta(x) = \,\frac{\alpha(x)-\alpha(0)}{\alpha(1)-\alpha(0)}\, , 
\end{equation}
where 
\begin{equation}
    \alpha(x) = \frac{1}{1+e^{s(h-x)}} \label{eq:logistic} 
\end{equation}
is the logistic function, $s$ is the steepness and $h$ is the inflection point. We can regain the NPD and VD by taking the limits $s\to 0$ and $s\to\infty$ respectively (see \Cref{fig:logistic_benefit}).
\subsection{The $\sigma$-rule: conditions for cooperation to be favoured}
\label{sec:favoured_coop}
For a particular update rule and population structure, the $\sigma$-rule allows us to determine which is the favoured strategy \cite{Wu2013}. We recall from \Cref{sec:conditions_for_success}, that a strategy $A$ is favoured over $B$, when $\rho_A>\rho_B$.

The $\sigma$-rule states that
\begin{equation}
    \rho_A>\rho_B \iff \sum_{j=0}^{k}\sigma_j (a_j-b_{k-j}) > 0 \, , \label{eq:sigma_rule}
\end{equation}
where $\sigma_j$ are the structure coefficients. It is assumed that the group size, $N=k+1$, is fixed, thus we have let $a_{j,k}=a_j$ and $b_{j,k}=b_j$. The structure coefficients are dependent on the population structure and update rule, but not the payoffs. Therefore if we calculate $\sigma_j$ for a given population structure and update rule, we can determine the favoured strategy for any game. 

For certain population structures, such as the well-mixed population and the cycle graph, the state is fully described by the number of $A$-players, $n$. Thus we can define the ratio of fixation probabilities as
\begin{equation}
    \frac{\rho_A}{\rho_B} = \prod_{n=1}^{Z-1}\frac{T_n^+}{T_n^-} \,, \label{eq:ratio_fixprobs}
\end{equation}
where $T_n^\pm$ are the transition probabilities to go from $n\to n\pm1$ $A$-type individuals \cite{Traulsen2009}. This does not hold in general, as the transition probabilities in more complex population structures will depend on the spatial configurations of different cell types, and thus are not uniquely defined by $n$. However, it is still possible to utilise this equation, as we see in \Cref{sec:global,sec:beneficial_coop}, by averaging over possible states to approximate $T_n^\pm$.
     
In the following we consider various cases where the structure coefficients can be calculated from transition probabilities in the weak selection limit, i.e.\ when $\delta \ll 1$. This limit is commonly employed within evolutionary graph theory in order to obtain analytical results, \mbox{e.g.\ }\mbox{\cite{Nathanson2009,Tarnita2009,Allen2016}}. Essentially, weak selection implies that the payoffs obtained by playing the game are only a small contribution to overall fitness. It allows expansion of fixation probabilities in powers of the selection strength parameter.

We outline known results for the well-mixed population, the cycle graph with death-birth update rule and $k$-regular graphs with death-birth update rule. We also introduce some new results, deriving the structure coefficients for the cycle graph with birth-death and shift update rules. Finally, we derive a new approximate expression for the structure coefficients of any population structure with global updating.

\subsubsection{Well-mixed population} \label{sec:well_mixed}
The structure coefficients for a well-mixed population are given by \cite{Gokhale2010}:
\begin{equation} \label{eq:sigma_wm}
    \sigma_j = \begin{cases}
        1\,, &\text{ if }0\le j \le N-2\\
        \frac{Z-N}{Z}\, , &\text{ if }j=N-1
    \end{cases}
\end{equation}
(see also \Cref{sec:global}). Thus we can obtain the condition for $\rho_A>\rho_B$ by plugging these into \Cref{eq:sigma_rule}. For a PGG defined by \Cref{eq:pgg} the condition that cooperators are favoured is
\begin{equation}
   \frac{Z-N}{Z}b\left[\beta(1)-\beta(0)\right] > \sum_{j=0}^{N-1}\sigma_j c \, .
\end{equation}
This becomes 
\begin{equation}
    \frac{b}{c}>\frac{N(Z-1)}{Z-N} \, , \label{eq:wm_coop_fav}
\end{equation}
when we set $\beta(1)=1$ and $\beta(0)=0$. Clearly the shape of the benefit function does not impact whether cooperation is favoured. For a large population this condition becomes $b/c>N$.
\subsubsection{Cycle graph: death-birth update}

We can obtain exact expressions for the structure coefficients of the cycle graph, in the weak selection limit. The cycle is a one-dimensional lattice with periodic boundary conditions. Individuals interact with their two nearest-neighbours, thus we have group size $N=3$.

The structure coefficients for the death-birth update rule are derived in \cite{Pena2016}. They are given by
\begin{align}
    \sigma_0 =1\,, && \sigma_1 = Z-2\,, && \sigma_2 = Z-3\,.
\end{align}
From \Cref{eq:sigma_rule} we obtain the condition for cooperation to be favoured under an NPD, defined by \Cref{eq:NPD}:
\begin{equation}
    \frac{b}{c}> \frac{3(Z-2)}{2(Z-3)}\,,
\end{equation}
which for $Z\to \infty$ becomes $b/c>3/2$. These conditions are lower than those obtained for a well-mixed population. For a general PGG defined by \Cref{eq:pgg} we can write down the condition
\begin{equation}
    \frac{b}{c} > \frac{2(Z-2)}{(Z-3)[\beta(1)+\beta(2/3)-\beta(1/3)-\beta(0)]}\, .
\end{equation}

\subsubsection{Cycle graph: birth-death update}
We derive novel results for the birth-death and shift update rules on the cycle, using a similar method to \mbox{\cite{Pena2016}} for the death-birth update rule. For the cycle, the transition probabilities are uniquely defined by the number of $A$-players in the population, $n$. Thus we can write down the ratio of transition probabilities for each $n$. For a birth-death update rule these are
\begin{equation}
    \frac{T_n^+}{T_n^-} = \begin{cases}
        (1+\delta a_0)/(1+\delta b_1)\,, &\text{ if } n=1 \\
        (1+\delta a_1)/(1+\delta b_1)\,, &\text{ if } 1<n<Z-1 \\
        (1+\delta a_1)/(1+\delta b_2)\,, &\text{ if } n=Z-1 \,.
    \end{cases}
\end{equation}
Substituting these into \Cref{eq:ratio_fixprobs}, and taking the limit, $\delta \ll 1$, we obtain
\begin{equation}
    \frac{\rho_A}{\rho_B} \approx 1+\delta[a_0-b_2+(Z-2)(a_1-b_1)]\, .
\end{equation}
In order that $\rho_A>\rho_B$, the second term must be positive. Thus, comparing this condition with \Cref{eq:sigma_rule}, we find the structure coefficients
\begin{align}
    \sigma_0 =1\,, &&\sigma_1 = Z-2\,, && \sigma_2= 0\,.
\end{align}

For the NPD, cooperation is favoured when
\begin{equation}
    \frac{b}{c}> \frac{3(Z-1)}{Z-3}\, ,
\end{equation}
which becomes $b/c>3$ in the large population limit, $Z\to\infty$. These conditions are equivalent to those obtained for the well-mixed population. For a general PGG defined by \Cref{eq:pgg} the condition is
\begin{equation}
    \frac{b}{c}> \frac{Z-1}{(Z-3)[\beta(2/3)-\beta(1/3)]} \, .
\end{equation}

\subsubsection{Cycle graph: shift update}
We follow the same procedure to derive the structure coefficients for the shift update rule. This time the ratio of transition probabilities is given by
\begin{equation}
    \frac{T_n^+}{T_n^-} = \begin{cases}
        \frac{(Z-1)(1+\delta a_0)}{2(1+\delta b_1)+(Z-3)(1+\delta b_0)} \,, &\text{ if } n=1 \\[5pt]
        \frac{(Z-n)(2(1+\delta a_1) +(n-2)(1+\delta a_2))}{n(2(1+\delta b_1)+(Z-n-2)(1+\delta b_0))} \,, &\text{ if } 1<n<Z-1 \\[5pt]
        \frac{2(1+\delta a_1)+(Z-3)(1+\delta a_2)}{(Z-1)(1+\delta b_2)} \,, &\text{ if } n=Z-1 \,.
    \end{cases}
\end{equation}
In the weak selection limit, $\delta\ll1$, \Cref{eq:ratio_fixprobs} becomes
\begin{equation}
    \frac{\rho_A}{\rho_B} \approx 1 + \delta \left[ (a_0-b_2) +2(H_{Z-1} -1)(a_1-b_1)) + (Z-2H_{Z-1})(a_2-b_0) \right] \, ,
\end{equation}
where $H_m$ is the $m$-th harmonic number:
\begin{equation}
    H_{m} = \sum_{n=1}^{m}\frac{1}{n} \, .
\end{equation}
Thus the structure coefficients are given by
\begin{align}
    \sigma_0 = 1\,, && \sigma_1 = 2(H_{Z-1}-1)\,, && \sigma_2 = (Z-2H_{Z-1})\,.
\end{align}

The condition for cooperation to be favoured in the NPD is
\begin{equation}
    \frac{b}{c}>\frac{3(Z-1)}{3(Z-1)-4H_{Z-1}} \,.
\end{equation}
In the large population limit this becomes $b/c>1$. As this condition is required in the definition of the NPD, we can state that cooperation is always favoured in the large population limit for a shift update under weak selection.

In fact, if we consider a general cooperation game as defined by \Cref{eq:pgg} we obtain the condition
\begin{equation}
    \frac{b}{c}> \frac{1}{\beta(1)-\beta(0)}
\end{equation}
in the large population limit, $Z\to\infty$. Letting $\beta(1)=1$ and $\beta(0)=0$, we regain the condition $b/c>1$. Thus for the shift update on the cycle, as with the well-mixed population, the condition for cooperation to be favoured is not dependent on the shape of the benefit function (although in this case we required the large population limit). Furthermore cooperation is favoured on the cycle with shift update for all PGGs, as defined by \Cref{eq:pgg}, given that the population is sufficiently large.
\subsubsection{Approximate results for $k$-regular graphs}
In higher dimensions the transition probabilities are no longer uniquely defined by the number of $A$-players in the population, but depend also on their configuration. Peña et al \cite{Pena2016} have derived expressions for the structure coefficients of regular graphs of degree $k\ge 3$, with death-birth updating, using pair approximation and diffusion approximation \cite{Ohtsuki2006}. They compared theoretical predictions with simulation results for the case of a volunteer's dilemma game. They find a good fit for random regular graphs, but that the approximations underestimate the critical benefit-to-cost ratio for lattices.

We do not state the full expressions here which are non-trivial functions of $k$. The condition for cooperation to be favoured with the NPD in the large population limit ($Z\gg k$) is given by \cite{Pena2016}
\begin{equation}
    \frac{b}{c}> \frac{k+1}{2}\, .
\end{equation}

\subsubsection{Structure coefficients under global updating}
\label{sec:global}

In the following we derive novel results for the structure coefficients under global updating. We find a general expression which is exact under certain conditions, and provides an approximation for the structure conditions for any population structure with global update rule. The proceeding sections have considered games played on a fixed graph or well-mixed population, within groups of fixed size, $N$. For well-mixed populations we were free to choose $N$ (although some results required $N\ll Z$), while for regular graphs we set $N=k+1$, where $k$ is the degree of the graph. Here we relax this condition and allow for variable group size. 

We make the assumption that there is a fixed distribution, $f_j^{A/B}(n,k)$, defining the probability that an $A/B$-player interacts with $j$ co-players of type $A$, given it has $k$ co-players in total and there are $n$ players of type $A$ in the population. If the population were defined on a graph, this would be the probability of an $A/B$-player having $j$ $A$-type neighbours, given $k$ total neighbours. This assumption is true for a well-mixed population or cycle graph, but not necessarily for other population structures where $f_j^{A/B}(n,k)$ depends on the specific configuration of players. The frequency of individuals with $k$ neighbours is given by $g_k$. We make the further assumptions that this distribution is fixed, and does not depend on type. See Appendix~\ref{sec:neighbour_dist_appendix} for a discussion of the validity of this assumption for the VT model.

In general, for a global update rule, we can define the transition probabilities
\begin{align}
    T_n^+ = \frac{Z-n}{Z}\frac{nF_A}{nF_A+(Z-n)F_B} && T_n^- = \frac{n}{Z}\frac{(Z-n)F_B}{nF_A+(Z-n)F_B} \, , \label{eq:transition_probs_global}
\end{align}
where
\begin{align}
    F_A &= 1 + \delta \sum_{k=1}^{Z-1}\sum_{j=0}^{k}f_j^A(n,k)g_ka_{j,k} \label{eq:VT_fitnessA} \\
    F_B &= 1 + \delta \sum_{k=1}^{Z-1}\sum_{j=0}^{k}f_j^B(n,k)g_kb_{j,k} \label{eq:VT_fitnessB} 
\end{align}
are the population averaged fitnesses. The payoffs $a_{j,k}$ and $b_{j,k}$ depend explicitly on the number of neighbours $k$.

Substituting \Cref{eq:transition_probs_global,eq:VT_fitnessA,eq:VT_fitnessB} into \Cref{eq:ratio_fixprobs}, and taking the weak selection limit we obtain
\begin{equation}
    \frac{\rho_A}{\rho_B} \approx 1+ \delta \underbrace{\sum_{n=1}^{Z-1}\sum_{k=1}^{Z-1}\sum_{j=0}^{k}g_k[f_j^A(n,k)a_{j,k}-f_j^B(n,k)b_{j,k}]}_\Gamma \, .
\end{equation}
Thus $\rho_A>\rho_B$ when $\Gamma>0$. In the weak selection limit,
\begin{equation}
    f_j^A\, (n,k) = f_{k-j}^B\,(Z-n,k) \, \label{eq:prob_identity}
\end{equation}
must hold by symmetry, and thus
\begin{equation}
    \sum_{n=1}^{Z-1}f_j^A(n,k) = \sum_{n=1}^{Z-1}f_{k-j}^B(n,k) \, .
\end{equation}
Therefore we have
\begin{align}
    \Gamma &= \sum_{k=1}^{Z-1}\sum_{j=0}^{k}\sum_{n=1}^{Z-1}g_kf_j^A(n,k)(a_{j,k}-b_{k-j,k}) \, .
\end{align}
The condition for \emph{A} to be favoured over \emph{B} is thus given by
\begin{equation}
    \sum_{k=1}^{Z-1}\sum_{j=0}^{k}\sigma_{j,k}(a_{j,k}-b_{k-j,k}) >0  \label{eq:sigma_rule_var}\,,
\end{equation}
where
\begin{equation}
    \sigma_{j,k} = g_k\sum_{n=1}^{Z-1}f_j^A(n,k) \label{eq:structure_coeffs_jk}
\end{equation}
are the structure coefficients. For a fixed group size, $N=k+1$, this reduces to \Cref{eq:sigma_rule}, with
\begin{equation}
    \sigma_j = \sum_{n=1}^{Z-1}f_j^A(n) \, , \label{eq:global_update_sigma}
\end{equation}
where we have dropped the explicit dependence on $k$.

Recall that this derivation is based on the assumption that $g_k$ and $f_j^A(n,k)$ are fixed. While this is not true in most cases, we can obtain an approximation for the structure coefficients by averaging over a large ensemble of population configurations, i.e.\ letting $f_j^A(n) = \langle f_j^A(n)\rangle_0$. Here, $\langle . \rangle$ represents the mean taken over possible configurations and the $0$ indicates that these are obtained in the neutral selection limit, i.e.\ $\delta=0$.

The well-mixed population is an example where $f_j^A(n)$ is fixed. It is defined by a hypergeometric distribution:
\begin{equation}
    f_j^A(n) = \binom{Z-1}{k}^{-1}\binom{n-1}{j}\binom{Z-n}{k-j} \, .
\end{equation}
We can therefore find the structure coefficients \cite{Gokhale2010} by substituting this expression for $f_j^A(n)$ into \Cref{eq:global_update_sigma}:
\begin{align}
    \sigma_j &= \binom{Z-1}{k}^{-1}\underbrace{\sum_{n=1}^{Z-1}\binom{n-1}{j}\binom{Z-n}{k-j}}_S \, .
\end{align}
It can be shown (see Appendix A in \mbox{\cite{Gokhale2010}}) that
\begin{equation} 
    S = 
\begin{cases}
    \binom{Z}{k+1} &\text{ if }0\le j<k \\[5pt]
    \binom{Z-1}{k+1} &\text{ if }j=k \,.
\end{cases}
\end{equation}
Thus the structure coefficients are given by
\begin{equation}
    \sigma_j = 
\begin{cases}
    \frac{Z}{k+1} &\text{ if }0\le j<k \\[5pt]
    \frac{Z-k-1}{k+1} &\text{ if }j=k \,.
\end{cases} \label{eq:sigma_wm2}
\end{equation}
These are equivalent to \Cref{eq:sigma_wm} up to a constant factor. The cycle graph also has a fixed distribution, $f_j^A(n)$, thus the structure coefficients for the shift update rule can also be obtained exactly using \Cref{eq:global_update_sigma}.

For a variable group size the structure coefficients for the well-mixed population are given by
\begin{equation}
    \sigma_{j,k} = g_k\sigma_j(k) = 
    g_k\begin{cases}
        \frac{Z}{k+1} &\text{ if }0\le j<k \\[5pt]
        \frac{Z-k-1}{k+1} &\text{ if }j=k,
    \end{cases}
\end{equation}
where $\sigma_j(k)$ are defined in \Cref{eq:sigma_wm2}.

As we have seen in previous sections, once the structure coefficients have been determined, we can use \Cref{eq:sigma_rule} or \Cref{eq:sigma_rule_var} to find the condition under which cooperation is favoured. For a PGG defined by \Cref{eq:pgg} this is given by
\begin{equation}
    \frac{b}{c} > \frac{Z-1}{\sum_{k=1}^{Z-1}\sum_{j=0}^k\sigma_{j,k}\left[\beta\left(\frac{j+1}{k+1}\right)-\beta\left(\frac{k-j}{{k+1}}\right)\right]} \, . \label{eq:global_coop_favoured}
\end{equation}

\subsection{Conditions for cooperation to be beneficial under global updating}
\label{sec:beneficial_coop}

Thus far, we have considered conditions under which a mutant is favoured. However, we recall from \mbox{\Cref{sec:conditions_for_success}}, that an alternative measure of mutant success can be obtained by considering the conditions under which it is beneficial. Here, we derive the condition for an $A$-mutant to be beneficial, \mbox{i.e.\ }$\rho_A>\rho_0$.

As in the previous section, we make the assumption that the distributions $g_k$ and $f_j^{A/B}(n,k)$ are fixed. Thus the population averaged fitnesses of $A$ and $B$ players are defined by \Cref{eq:VT_fitnessA,eq:VT_fitnessB} and the transition probabilities by \Cref{eq:transition_probs_global}. The fixation probability for a single $A$-mutant \cite{Traulsen2009} is then given by 
\begin{equation}
    \rho_A = \left[1+ \sum_{m=1}^{Z-1}\prod_{n=1}^{m}\frac{T_n^-}{T_n^+}\right]^{-1} \, .
\end{equation}
Substituting in the transition probabilities and taking the weak selection limit $\delta \ll 1$ we obtain
\begin{equation}
    \rho_A = \frac{1}{Z} + \frac{\delta}{Z^2} \sum_{k=1}^{Z-1}\sum_{j=0}^k \left(\theta^A_{j,k}a_{j,k}-\theta^B_{j,k}b_{j,k}\right) +\mathcal{O}(\delta^2)\, ,
\end{equation}
where we have defined
\begin{align}
    \theta^A_{j,k} &= g_k\sum_{m=1}^{Z-1}\sum_{n=1}^{m}f_j^A(n,k)  \label{eq:theta_A} \\
    \theta^B_{j,k} &= g_k\sum_{m=1}^{Z-1}\sum_{n=1}^{m}f_j^B(n,k) = g_k\sum_{m=1}^{Z-1}\sum_{n=1}^{m}f^A_{k-j}(Z-n,k)\label{eq:theta_B} \, .
\end{align}
The final equality is obtained by symmetry arguments in the weak selection limit. 

The condition for $A$ to be a beneficial mutation, $\rho_A>1/Z$, is therefore given by
\begin{equation}
    \sum_{k=1}^{Z-1}\sum_{j=0}^k \left(\theta^A_{j,k}a_{j,k}-\theta^B_{j,k}b_{j,k}\right) > 0 \, .
\end{equation}
If we consider a PGG as defined by \Cref{eq:pgg}, then cooperation is beneficial when
\begin{equation}
    \frac{b}{c} > \frac{Z(Z-1)}{2\sum_{k=1}^{Z-1}\sum_{j=0}^{k}\left[\theta^A_{j,k}\,\beta\left(\frac{j+1}{k+1}\right)-\theta^B_{j,k}\,\beta\left(\frac{j}{k+1}\right)\right]} \, . \label{eq:global_coop_beneficial}
\end{equation}
For a fixed group size $N=k+1$ these conditions simplify to
\begin{equation}
    \sum_{j=0}^k \left(\theta^A_{j}a_{j}-\theta^B_{j}b_{j}\right) > 0 
    \label{eq:beneficial_cooperation_fixed_k}
\end{equation}
and
\begin{equation}
      \frac{b}{c} > \frac{Z(Z-1)}{2\sum_{j=0}^{k}\left[\theta^A_{j}\,\beta\left(\frac{j+1}{k+1}\right)-\theta^B_{j}\,\beta\left(\frac{j}{k+1}\right)\right]} \, ,
\end{equation}
where 
\begin{equation}
    \theta_j^{A/B} = \sum_{m=1}^{Z-1}\sum_{n=1}^mf_j^{A/B}(n,k) \, .
    \label{eq:theta_fixed_k}
\end{equation}
\section{Public goods games in an epithelium}
\label{sec:epithelium}
A number of studies have considered the evolutionary dynamics of sigmoid PGGs in epithelia, representing the tissue either as a well-mixed population \cite{Archetti2013b}, or a fixed graph structure with various local update rules \cite{Archetti2013a,Archetti2016}. Here we use the framework introduced in \cite{Renton2019} to incorporate explicit tissue dynamics, using the Voronoi tessellation (VT) model, with a spatially decoupled (global) update rule. This means that when the population is updated, a division and death occur simultaneously, but there is no spatial dependence between the two events. 

In this section we briefly introduce the VT model, before calculating conditions under which cooperation is favoured and beneficial for a sigmoid PGG. We verify theoretical results by running simulations in various parameter regimes. We also compute the gradient of selection in order to obtain a fuller picture of the dynamics. In all cases we compare VT model results with the well-mixed population. 

\subsection{Voronoi tessellation model}

The VT model represents a tissue as a set of points, corresponding to cell centres \cite{Meineke2001,VanLeeuwen2009}. The shape of each cell, as well as its neighbour connections, is determined by performing a Voronoi tessellation. Cells move subject to spring-like forces, which they exert on their neighbours.

The population evolves through a process of sequential update events, each consisting of a cell division and a cell death, which occur simultaneously. We choose to temporally couple division and death in this way to maintain a constant population size. Allowing separate stochastic birth and death processes, without some other mechanism to maintain homeostatic population size, would result in population extinction or rapid growth. This is something we will address in future work, by introducing contact inhibition as a means of controlling the population size.

Update events occur at rate $\lambda$, according to a continuous time Moran process. When an update occurs, a cell is chosen to divide with probability proportional to fitness. This parent cell is removed from the tissue and replaced with two identical progeny cells, separated by a distance $\epsilon$, across a uniformly random axis. Simultaneously, a cell is chosen to die uniformly at random, and is removed from the tissue. A full description of the VT model used is given in Appendix~\ref{sec:VT_appendix}. 

\begin{figure}[h!t]
    \centering
    \includegraphics{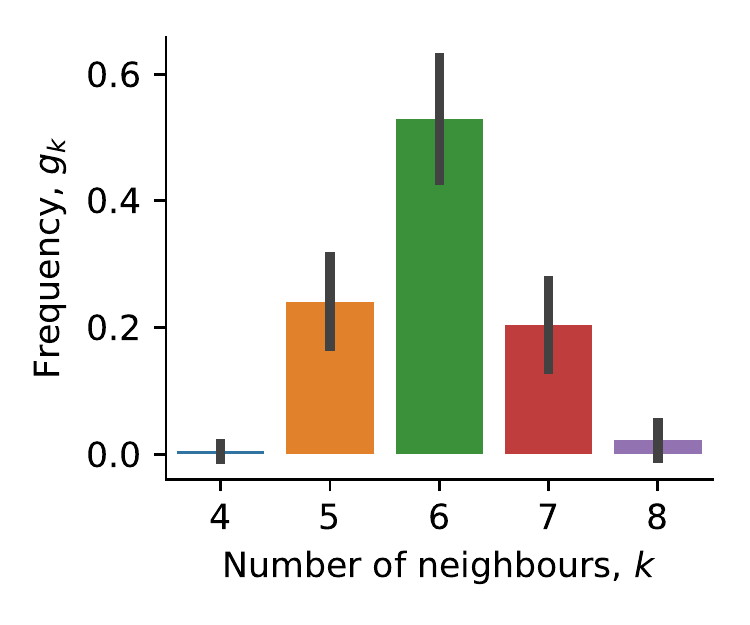}
    \caption{Degree distribution for the Voronoi tessellation model. Error bars show standard deviation. Data is obtained from simulations with population size, $Z=100$.}
    \label{fig:degree_dist}
\end{figure}

\begin{figure}[h!t]
    \centering
    
    \includegraphics{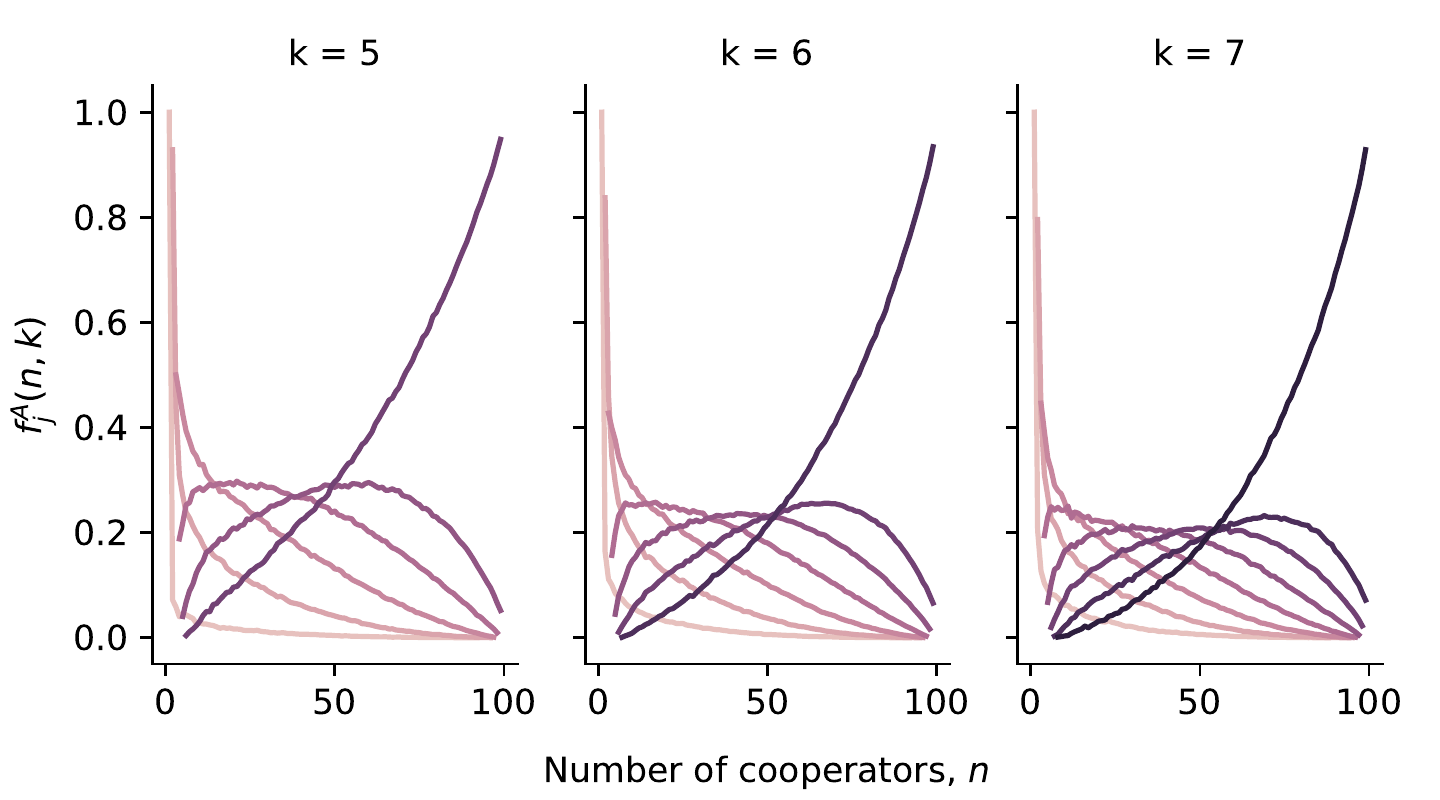}
    \includegraphics{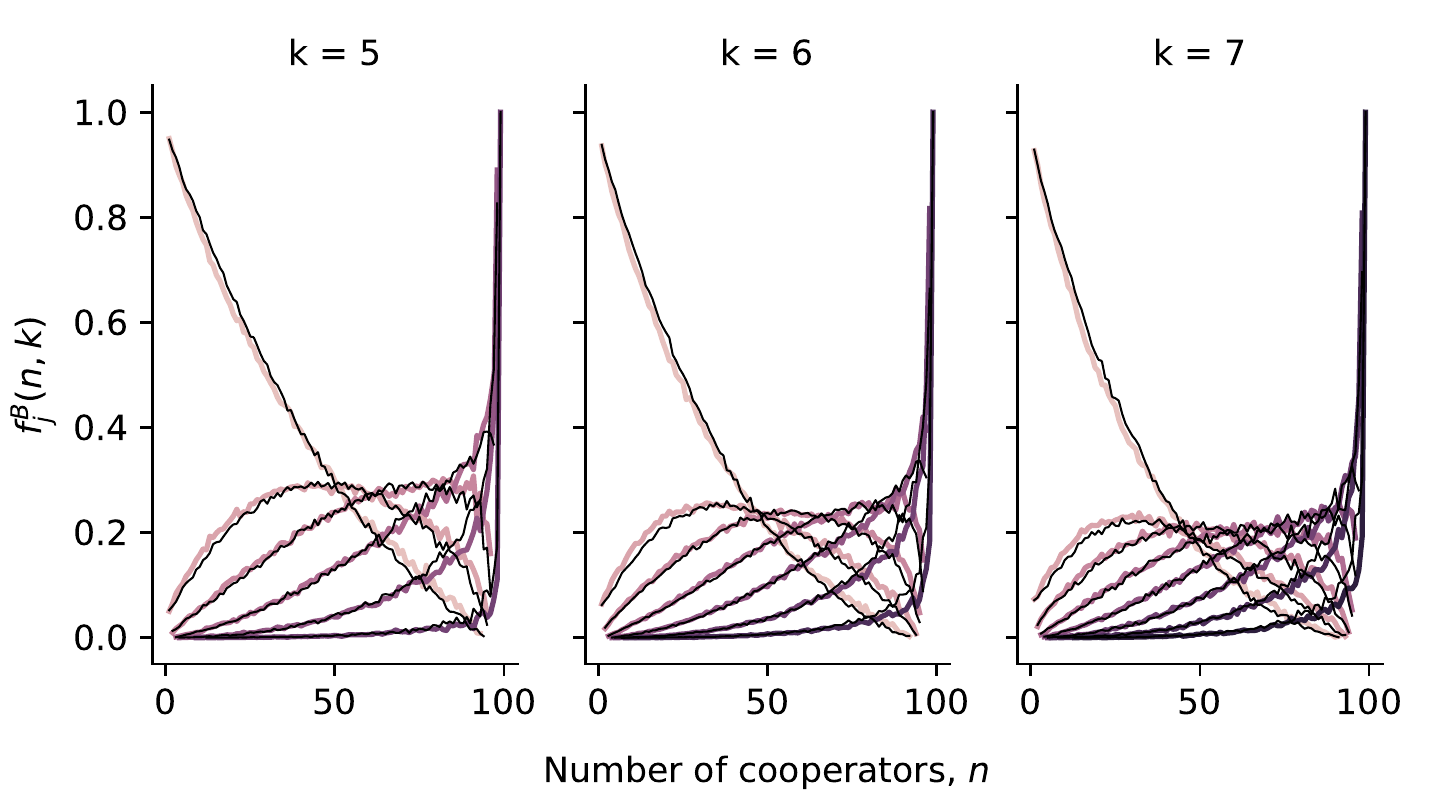}\\[0.5cm]
    \qquad\includegraphics[scale=1.0]{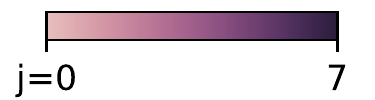}
    \caption{Frequency distributions $f_j^A(n,k)$ and $f_j^B(n,k)$ for $Z=100$. These define the probability that a cell of type $A/B$ has $j$ neighbours of type $A$, given $k$ neighbours total and $n$ cells of type $A$ in the population.  The lower panel compares values of $f_j^B(n,k)$ calculated directly through simulation (black) with values obtained from the simulated data for $A$ cells defined by $f_j^B(n,k)=f_{k-j}^A(Z-n,k)$.}
    \label{fig:A-dist}
\end{figure}

We obtain $g_k$ and $f_j^A(n,k)$ by averaging over a large ensemble of possible states in the weak selection limit. We then make the assumption that variation around this mean can be neglected. \Cref{fig:degree_dist,fig:A-dist} show the distributions $g_k$ and $f_j^A(n,k)$ for the VT model under neutral selection, calculated by averaging over $500$ simulations, each of which starts with a single neutral mutant and is run to fixation. An example simulation is shown in  \mbox{\Cref{fig:example_VT_simulations}}. See Appendix~\ref{sec:neighbour_dist_appendix} for further discussion on neighbour distributions in the VT model and the validity of assuming $g_k$ is independent of $n$ and cell type.

\begin{figure}[h!tbp]
    \centering
    \includegraphics{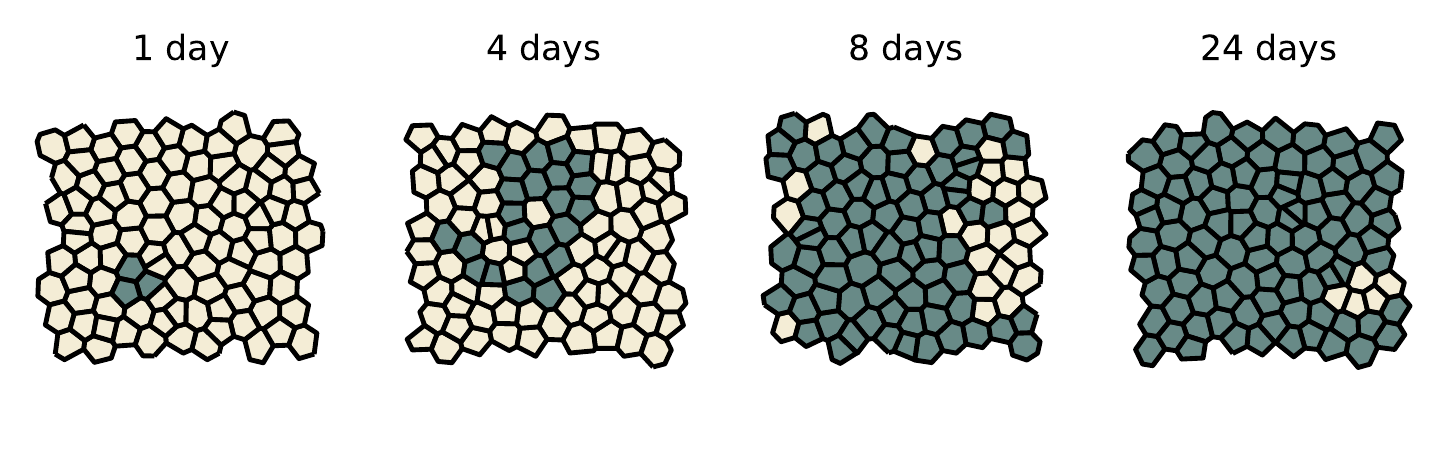}
    \caption{Time snapshots for a simulation of mutant invasion in the Voronoi tessellation model with decoupled update rule. The simulation is initialised with a single neutral mutant (grey) in a population of $Z=100$ cells and run until fixation. Selection is neutral ($\delta=0$), so all cells have equal fitness. Parameters for the Voronoi tessellation model are given in \Cref{table:params}. }
    \label{fig:example_VT_simulations}
\end{figure}


\subsection{Favourable cooperation}

\begin{figure}[h!]
    \centering
    \includegraphics{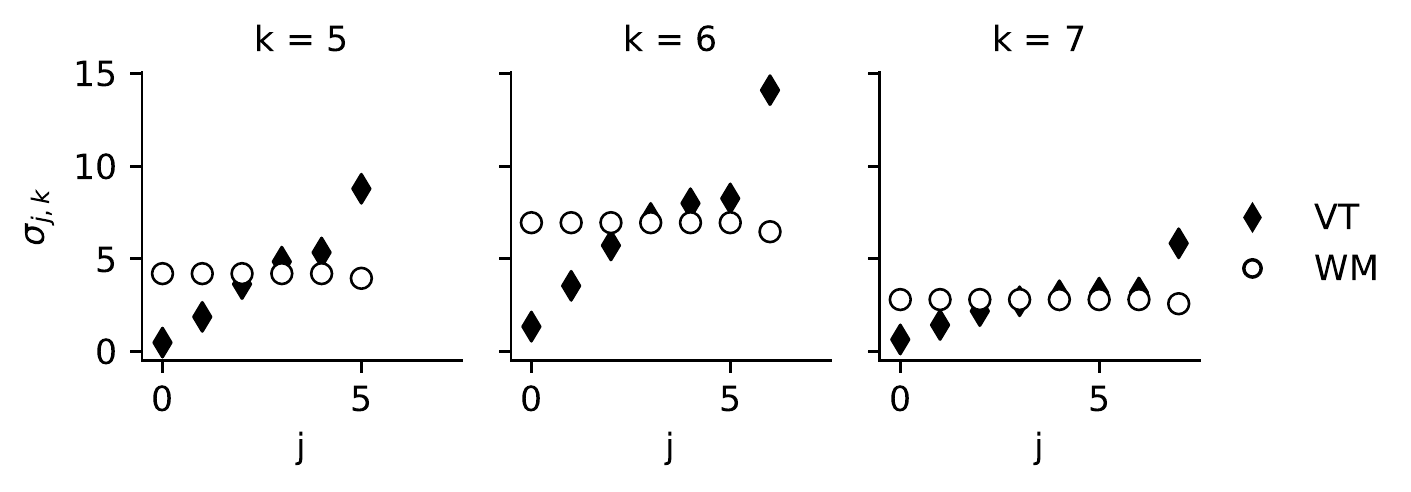}
    \caption{Comparing the structure coefficients for the Voronoi tessellation model with decoupled update (VT) and a well-mixed (WM) population with variable group size. Variation in group size arises naturally in the VT model due to its neighbour distribution, which is plotted in \Cref{fig:degree_dist}. We set the group size distribution for the WM population to be equal to that of the VT model. Members of each group are then selected uniformly at random for the WM population. }
    \label{fig:sigma}
\end{figure}

The condition for cooperation to be favoured can be approximated by calculating the structure coefficients using \Cref{eq:structure_coeffs_jk}. \Cref{fig:sigma} plots the VT structure coefficients with those for a well-mixed population as defined by \Cref{eq:sigma_wm2}.

Using the structure coefficients we can derive the condition for cooperation to be favoured for an arbitrary PGG, as defined by \Cref{eq:pgg}. We define the critical benefit-to-cost ratio $(b/c)^*_1$, such that $\rho_C>\rho_D$ when $b/c>(b/c)^*_1$. Thus from \Cref{eq:global_coop_favoured} we can write
\begin{equation}
    \left(\frac{b}{c}\right)^*_1 = \frac{Z-1}{\sum_{k=1}^{Z-1}\sum_{j=0}^{k}
                                    \sigma_{j,k} \left[\beta\left(\frac{j+1}{k+1} \right) 
                                                -\beta\left(\frac{k-j}{k+1} \right)\right]} \, . \label{eq:btoc1}
\end{equation}
For an NPD, defined by \Cref{eq:NPD}, this becomes 
\begin{equation} \left(\frac{b}{c}\right)^*_1=\frac{Z-1}{\sum_{k=1}^{Z-1}\sum_{j=0}^k\sigma_{j,k}\frac{2j+1-k}{k+1}} \, .
\end{equation}

Substituting in the structure coefficients we obtain $(b/c)_1^*\approx 2.22$ for the VT model with decoupled update rule and population size $Z=100$. For a well-mixed population with the same group size distribution we obtain $(b/c)_1^*\approx 7.35$. As we would expect there is a significant increase in the success of cooperative mutants under the VT model. This is due to the high level of assortment in the VT model, which means cooperators are likely to have more cooperator neighbours than defectors.

On average, cells have six neighbours, thus the mean group size is seven. We can therefore compare the critical benefit-to-cost ratio for a well-mixed population with variable group size, given above, to that of a well-mixed population with fixed group size, $N=7$. The latter is given by \Cref{eq:wm_coop_fav} to be $(b/c)^*_1=7.45$. Clearly, incorporating variation in group size into the well-mixed population has a negligible impact on whether cooperation is favoured. We note however, that the level of variation in group size we have considered, which is realistic for an epithelium, is small. Larger variation in group size, such as that observed for scale-free networks, may have a larger effect \mbox{\cite{Archetti2016}}.

\begin{figure}[htb]
    \centering
    \includegraphics{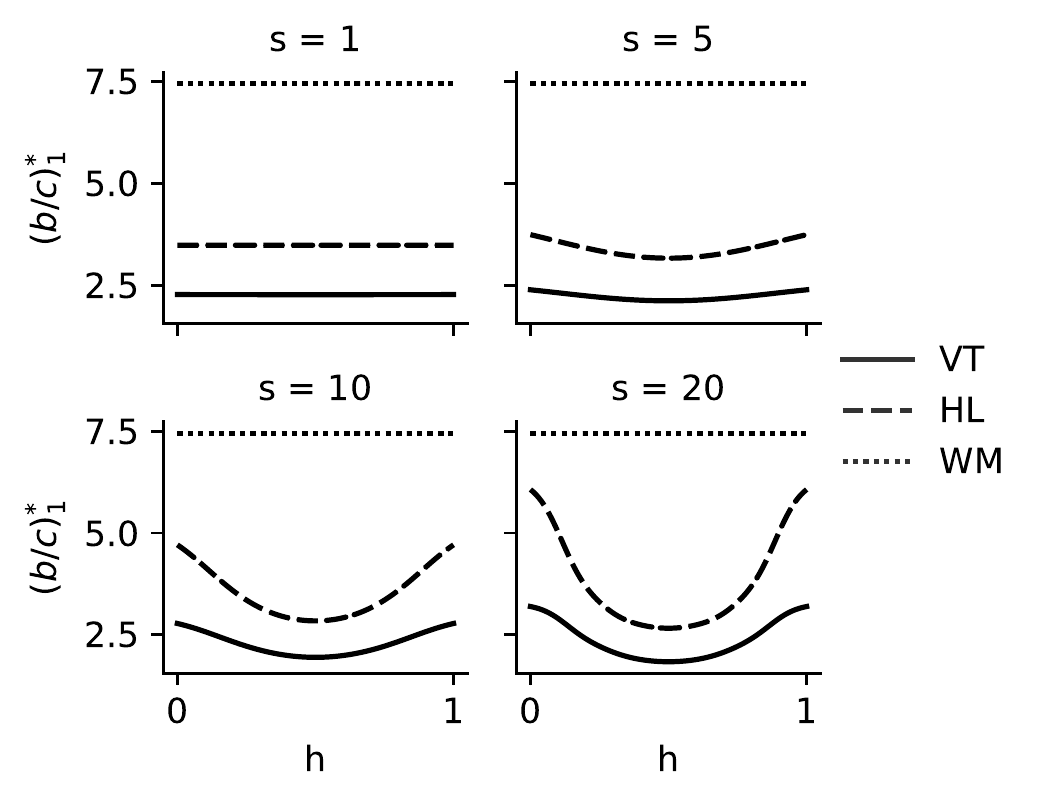}
    \caption{Comparing the critical benefit-to-cost ratio, $(b/c)^*_1$ at which $\rho_C>\rho_D$, for a logistic benefit function. For a well-mixed population with $N=7$ (WM), $(b/c)^*_1$ is highest, and independent of the inflection point, $h$, and steepness, $s$. For the Voronoi tessellation model with decoupled update (VT) and fixed hexagonal lattice with death-birth update (HL), $(b/c)^*_1$ varies with $h$ and $s$. For small $s$ the benefit function approaches linearity and we regain the results for an NPD.} \label{fig:compare_btoc1}
\end{figure}

We can also use \Cref{eq:btoc1} to determine $(b/c)^*_1$ for a sigmoid benefit function, defined by \Cref{eq:sigmoid_benefit}. Recall that the logistic function has two parameters: the steepness, $s$, and the inflection point, $h$. \Cref{fig:compare_btoc1} compares the predicted values of $(b/c)^*_1$ for the VT model, with those for a well-mixed (WM) population with group size 7, and hexagonal lattice (HL) with death-birth update rule. These are obtained from \Cref{eq:btoc1} by using the relevant structure coefficients in each case (structure coefficients for death-birth update on regular graphs are derived in \cite{Pena2016}). 

Values of $(b/c)^*_1$ are symmetric across $h=0.5$ for all three cases, and minimised at $h=0.5$ for the hexagonal lattice and VT model. In Appendix~\ref{sec:minimising_btoc_appendix} we show that $(b/c)^*_1$ is in fact minimised at $h=0.5$, so long as the structure coefficients increase with $j$ for $0\le j<k$. For the well-mixed population $(b/c)^*_1$ does not vary with either $s$ or $h$. Furthermore, it is clear for all population types, that as the NPD is approached ($s\to 0$), $(b/c)^*_1$ becomes independent of $h$.

In all parameter regimes, $(b/c)^*_1$ is highest for the well-mixed population. Both the VT model with decoupled update and HL with death-birth update show similar variation with $s$ and $h$, however $(b/c)^*_1$ is always lower for the VT model. Therefore in terms of thresholds for favourability, we can determine that cooperation is most successful in the VT model with decoupled update, followed by the hexagonal lattice with death-birth update. Cooperation does least well in the well-mixed population. This suggests that both population structure and global updating promote cooperation. 

\begin{figure}[htb]
    \centering
    \includegraphics{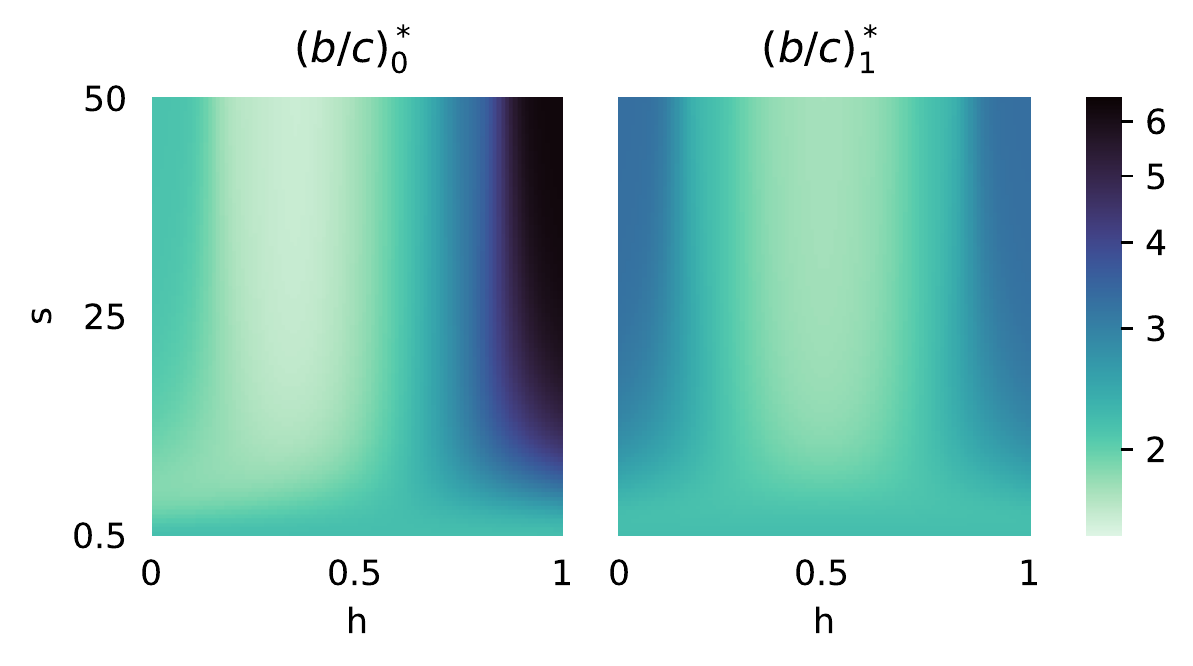}
    \caption{Critical benefit-to-cost ratios for the VT model with decoupled update. These are given by \Cref{eq:btoc0,eq:btoc1} for a PGG with logistic benefit function, defined by \Cref{eq:sigmoid_benefit}. Parameters $s$ and $h$ correspond to the steepness and inflection point of the benefit function, respectively. Cooperation is beneficial when $b/c>(b/c)^*_0$ (left) and favoured when $b/c>(b/c)^*_1$ (right).}
    \label{fig:sigmoid_critical}
\end{figure}

\begin{figure}[htb!]
    \centering
    \includegraphics{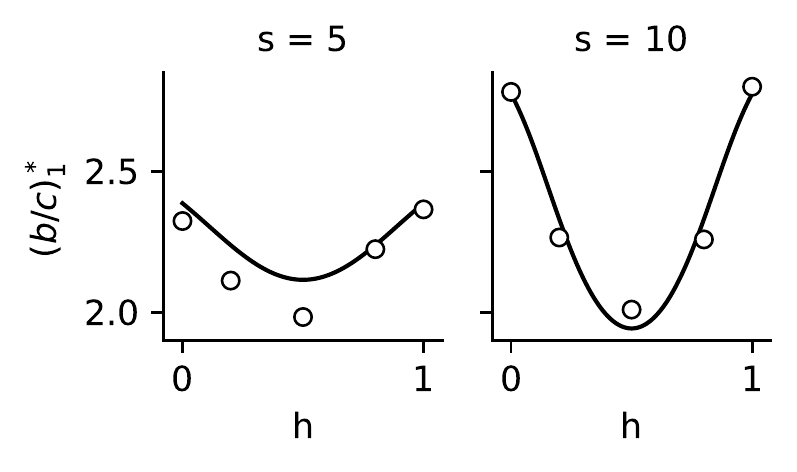}
    \caption{Critical benefit-to-cost ratio, $(b/c)^*_1$, above which $\rho_C>\rho_D$, for a logistic benefit function. The solid line plots \Cref{eq:btoc1} and circles are simulation data. For both $s=5$ and $s=10$ there is symmetry across $h=0.5$, at which point $(b/c)^*_1$ is minimised.} \label{fig:btoc1_sim}
\end{figure}

\Cref{fig:sigmoid_critical} (right panel) shows the variation of $(b/c)^*_1$ with $h$ and $s$ for the VT model. As we have discussed, these results are based on the approximation that $f^A_j(n,k)$ and $g_k$ are fixed.
In order to verify the accuracy of this approximation we compare \Cref{eq:btoc1} with simulation results in \Cref{fig:btoc1_sim}. Simulated values of $(b/c)^*_1$ were obtained for each parameter set $(s,h)$ as follows. We calculated $\rho_{C/D}$ for various $b/c$ values, by running $10^4$ simulations of the VT model to fixation, starting with a single $C/D$ mutant and population size $Z=100$. In all simulations we use small selection strength ($\delta=0.025$) and set $c=1$. Thus $(b/c)^*_1$ is determined by the point at which $\rho_C=\rho_D$.

There is a decent fit between simulation and theory. It is possible this could be improved by running larger numbers of simulations, however the model is computationally expensive. In any case the qualitative behaviour is consistent. For a fixed steepness, $s$, $(b/c)^*_1$ is minimised at $h=0.5$ and (near) symmetric across this value. The values of $(b/c)^*_1$ are highest when $h=0$ and $h=1$, where the benefit function provides diminishing returns or increasing returns respectively. 

\subsection{Beneficial cooperation}
\label{sec:epithelium_beneficial_coop}

Thus far we have considered conditions for cooperation to be favoured, i.e.\ where $\rho_C>\rho_D$. We can also define the critical benefit-to-cost ratio $(b/c)^*_0$ above which cooperation is beneficial, i.e.\ $\rho_C>\rho_0$. From \Cref{eq:global_coop_beneficial} this is given by
\begin{equation}
    \left(\frac{b}{c}\right)^*_0 = \frac{Z(Z-1)}{2\sum_{k=1}^{Z-1}\sum_{j=0}^{k}\left[\theta^A_{j,k}\,\beta\left(\frac{j+1}{k+1}\right)-\theta^B_{j,k}\,\beta\left(\frac{j}{k+1}\right)\right]} \, , \label{eq:btoc0}
\end{equation}
where $\theta^{A/B}_{j,k}$ are calculated from the distributions $f_j^{A/B}(n,k)$ and $g_k$ according to \Cref{eq:theta_A,eq:theta_B}.

\begin{figure}[htb!]
    \centering
    \includegraphics{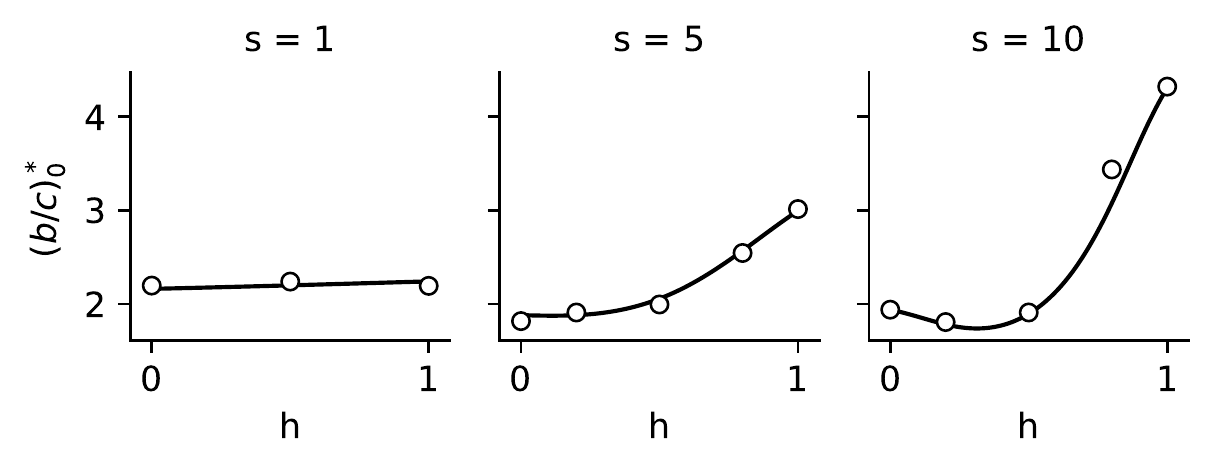}
    \caption{Critical benefit-to-cost ratio, $(b/c)^*_0$, above which $\rho_C>\rho_0$, for a logistic benefit function. The solid line plots \Cref{eq:btoc0} and circles are simulation data. For small $s$ the logistic benefit function becomes near linear and the game approaches an $NPD$, thus there is little variation in $(b/c)^*_0$. For larger $s$ there is strong dependence on the inflection point, $h$, particularly for $h>0.5$. } \label{fig:btoc0_sim}   
\end{figure}

\Cref{fig:sigmoid_critical} (left panel) plots $(b/c)^*_0$ against $s$ and $h$. We can see that for large $s$, $(b/c)^*_0$ is maximised at $h=1$ and has a minimum at $h\approx 0.35$. For smaller $s$ this minimum moves towards $h=0$. As $s$ decreases further, the logistic function approaches linear and there is negligible variation in $(b/c)^*_0$ with $h$. In the limit $s\to0$ the game becomes an NPD, with $(b/c)^*_0=(b/c)^*_1 \approx 2.2$. \Cref{fig:btoc0_sim} compares simulated values of $(b/c)^*_0$ with the theoretical prediction, finding good agreement between the two for a range of $s$ and $h$ values.

\begin{figure}[h!]
    \centering
    \includegraphics{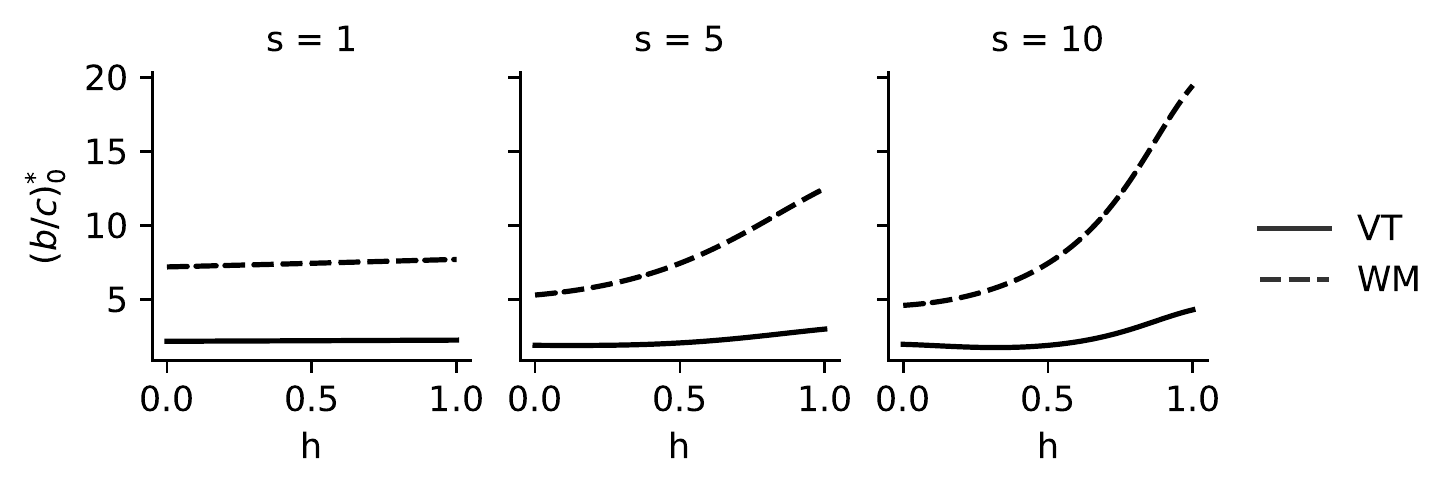}
    \caption{Comparing the critical benefit-to-cost ratio, $(b/c)^*_0$ at which $\rho_C>\rho_0$, for a PGG with logistic benefit function. The critical ratio is always higher for the well-mixed population with $N=7$ (WM), than for the Voronoi tessellation model with decoupled update (VT). For small $s$ the benefit function becomes near linear and variation of $(b/c)^*_0$ with $h$ is small. For WM, $(b/c)^*_0$ increases with $h$, taking its minimum value at $h=0$. By contrast, for VT, there is a minimum of $(b/c)^*_0$ at $h\approx 0.35$ when $s$ is sufficiently large. For both WM and VT, $(b/c)^*_0$ is maximised at $h=1$, for any given $s$.} \label{fig:compare_btoc0}
\end{figure}

We saw in \Cref{fig:compare_btoc1} that the critical benefit-to-cost ratios for cooperation to be favoured, $(b/c)^*_1,$ are lower in the VT model compared to the well-mixed population. \Cref{fig:compare_btoc0} plots $(b/c)^*_0$ for a well-mixed population with $N=7$ and the VT model with decoupled update, showing clearly that the critical benefit-to-cost ratios for cooperation to be beneficial are also lower for the VT model. 
Thus under both measures, cooperation is promoted by the VT model. In contrast to $(b/c)^*_1$, which was independent of the shape of the benefit function for the well-mixed population, $(b/c)^*_0$ is an increasing function of $h$, so long as $s$ is sufficiently large.

\begin{figure}[htb!]
    \centering
    \begin{subfigure}{0.49\textwidth}
        \includegraphics[scale=0.9]{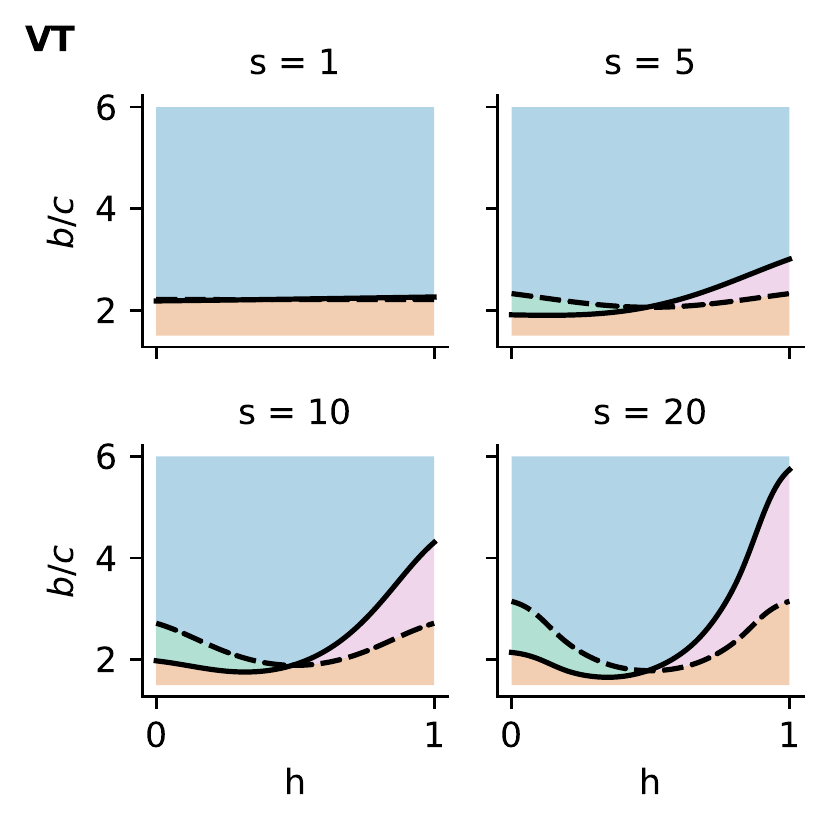}
    \end{subfigure}
    \begin{subfigure}{0.49\textwidth}
        \includegraphics[scale=0.9]{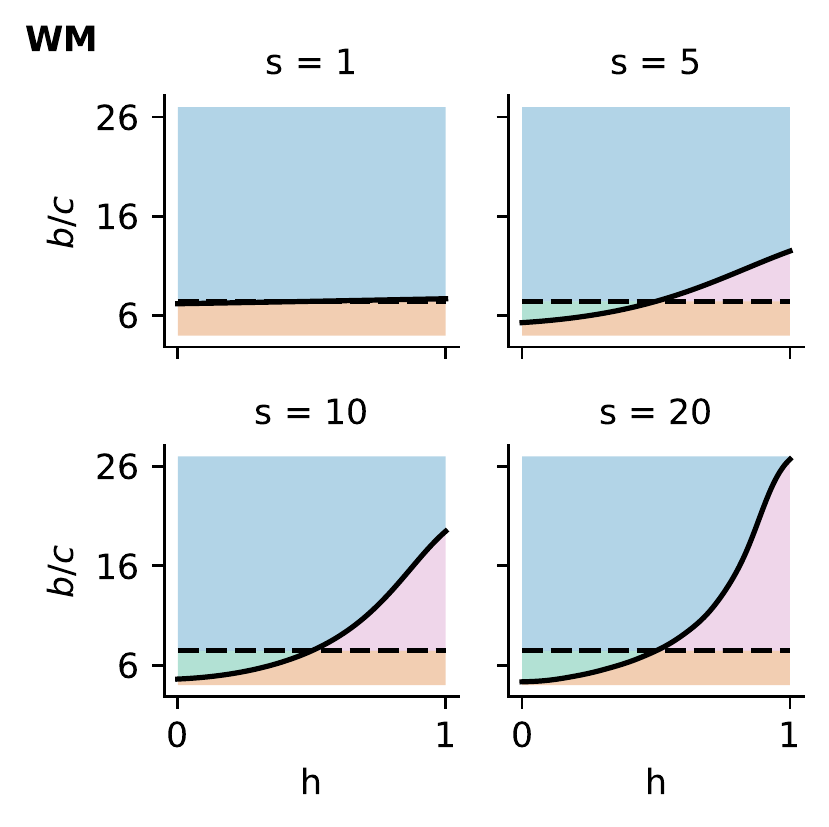}
    \end{subfigure}
    \caption{Success of cooperator mutants in the VT model (left) and well-mixed population (right), for a PGG with logistic benefit function. The solid line corresponds to $(b/c)^*_0$, where $\rho_C=\rho_0$. The dashed line corresponds to $(b/c)^*_1$, where $\rho_C=\rho_D$. \emph{Blue region (top):} $C$ is beneficial and favoured ($\rho_C>\rho_D$ and $\rho_C>\rho_0$). \emph{Green region (left):} $C$ is beneficial but not favoured ($\rho_D>\rho_C>\rho_0$). \emph{Pink region (right):} $C$ is favoured but not beneficial ($\rho_0>\rho_C>\rho_D$). \emph{Orange region (bottom):} $C$ is neither beneficial not favoured ($\rho_C<\rho_D$ and $\rho_C<\rho_0$).}
    \label{fig:coop_success}
\end{figure}

In general, conditions for cooperation to be beneficial are not equivalent to conditions for cooperation to be favoured. This is evident from \Cref{fig:coop_success}, where we plot $(b/c)^*_0$ and $(b/c)^*_1$ against $h$ for various values of $s$. The parameter space can be divided into regions where cooperation is both favoured and beneficial, favoured but not beneficial, beneficial but not favoured, and neither favoured nor beneficial. 

From \Cref{fig:coop_success} we can see that $(b/c)^*_0=(b/c)^*_1$ when $h=0.5$. Furthermore, as $s$ decreases, the regions of parameter space where cooperation is beneficial, but not favoured, or favoured, but not beneficial, get smaller. For sufficiently small $s$ we obtain $(b/c)^*_0\approx(b/c)^*_1$. In \mbox{Appendix~\ref{sec:equivalence_beneficial_favoured}} we show that the sigmoid public goods game satisfies a property called \emph{antisymmetry-of-invasion} when $s\to 0$ or $h=0.5$. This guarantees that the conditions for a mutant to be beneficial and favoured are equivalent. For both the VT model and well-mixed populations it is clear that behaviour where cooperation is beneficial but not favoured, is only possible when $h<0.5$. Conversely behaviour where cooperation is favoured but not beneficial occurs only when $h>0.5$.

We can understand this intuitively by considering the extreme cases ($h=0,1$) of the VD game, obtained by letting $s\to \infty$. When $h=0$, a cooperator always receives the full benefit, even if it has no cooperator neighbours. Defectors require a single cooperator neighbour to obtain the benefit. Thus both cooperators and defectors have higher than average fitness early on in the invasion process, when they are most vulnerable to extinction. It is therefore possible, depending on the benefit-to-cost ratio, that both perform better than a neutral invader, and therefore both are beneficial mutations. However, one can still be favoured over the other if its fixation probability is higher.

The converse is true when $h=1$: defectors will never receive any benefit, and cooperators only obtain the benefit when surrounded by other cooperators. Thus when the number of cooperators/defectors is small, they have lower than average fitness, and there is a high chance they die out early in the invasion process. Therefore, it is possible that neither performs better than a neutral invader.

\subsection{Gradient of selection}

We can obtain more insight into what is happening in the different parameter regions by looking at the gradient of selection, $G(n)=T^+(n)-T^-(n)$. The transition probabilities are defined by \Cref{eq:transition_probs_global}, thus in the weak selection limit, $\delta\ll 1$, the gradient of selection becomes
\begin{equation}
    G(n) \approx \frac{Z-n}{Z}\frac{n}{Z}\delta\left\{\sum_{k=1}^{Z-1}\sum_{j=0}^k g_k (f_j^A(n,k)a_{j,k}-f_j^B(n,k)b_{j,k})\right\} \, . \label{eq:gos}
\end{equation}
The sum essentially gives the difference in expected payoffs of $A$ and $B$ players. Thus, the right-hand side is identical to the replicator equation which describes the deterministic dynamics which can be obtained in the large-population limit. The sign of $G(n)$ indicates the direction of selection, and we can consider the roots of $G(n)$ as `fixed points'. Of course, for a finite population there are only two absorbing states, $n=0$ and $n=Z$, however the location of fixed points is still important. In particular the system may remain for a long time near a stable fixed point. We can classify the behaviour of the system in different parameter regions based on the fixed points of the gradient of selection.

\begin{figure}[p]
    \centering
    \begin{subfigure}{0.94\textwidth}
        \centering
        \includegraphics[scale=0.9]{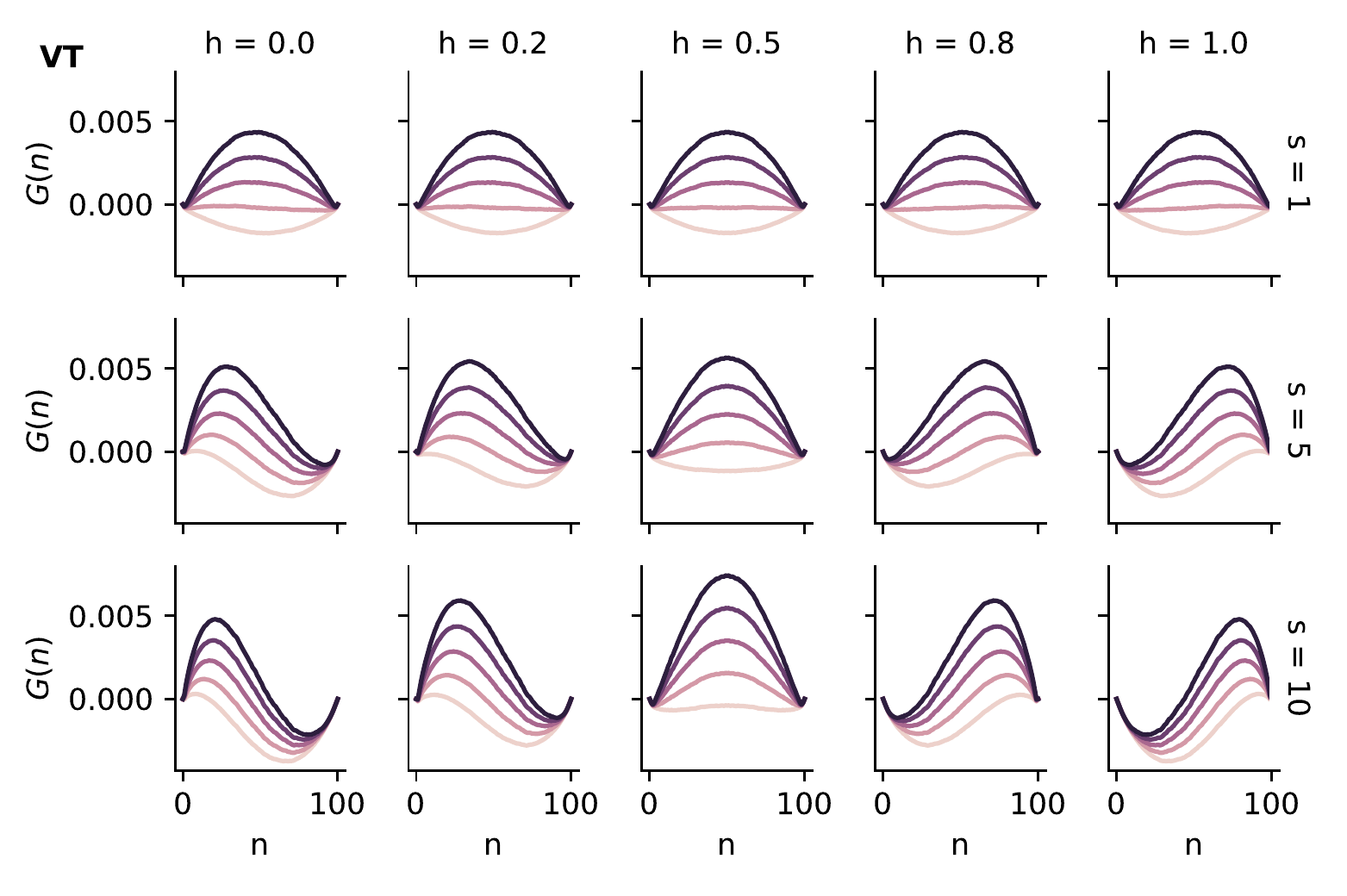}
    \end{subfigure}
    \begin{subfigure}{0.05\textwidth}
        \centering
        \includegraphics[scale=0.9]{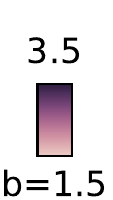}
    \end{subfigure}
    \begin{subfigure}{0.94\textwidth}
        \centering
        \includegraphics[scale=0.9]{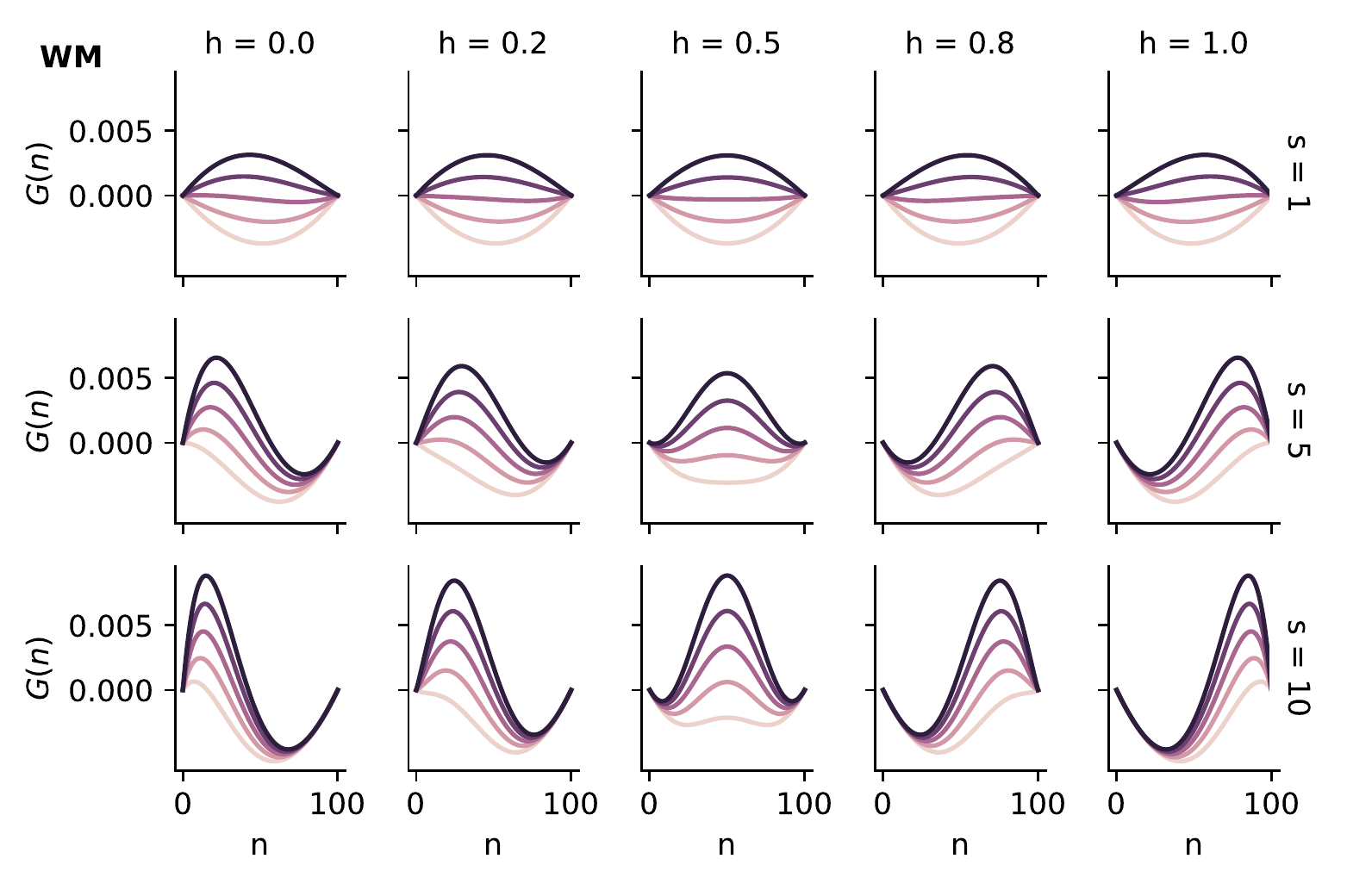}
    \end{subfigure}
    \begin{subfigure}{0.05\textwidth}
        \centering
        \includegraphics[scale=0.9]{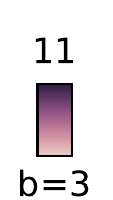}
    \end{subfigure} 
    \caption{Gradient of selection, $G(n)$ for a PGG with logistic benefit function in a well-mixed population (WM) and Voronoi tessellation model with decoupled update (VT). The qualitative behaviour is very similar between the two, however occurs at different values of the benefit, $b/c$. Where $G(n)>0$, selection is working to increase $n$, and vice versa. The roots of $G(n)=0$ can be considered as fixed points, and we can use these to classify the behaviour in different parameter regions. \label{fig:gos}}   
\end{figure}

\Cref{fig:gos} plots $G(n)$ for a PGG with various values of $h$, $s$ and $b/c$, both for the VT model and well-mixed population. There are four dynamical regimes, consistent with the deterministic results for PGG in a well-mixed population in \cite{Archetti2013b}:
\begin{enumerate}[(i)]
    \item \emph{Dominance:} there are only two fixed points at $n=0$ and $n=Z$. Defection dominates if the $n=0$ fixed point is stable, while cooperation dominates if the $n=Z$ fixed point is stable.
    \item \emph{Coexistence:} there is an internal stable fixed point, $n_R$, along with two unstable fixed points at $n=0$ and $n=Z$. Selection pushes the system towards the stable fixed point, thus it can take a long time to reach one of the absorbing states.
    \item \emph{Coordination:} there is an internal unstable fixed point, $n_L$, along with two stable fixed points at $n=0$ and $n=Z$.
    \item \emph{Coexistence \& coordination:} In addition to the fixed points at $n=0$  (stable) and $n=Z$ (unstable), there is both an unstable internal fixed point on the left, $n_L$, and a stable internal fixed point on the right, $n_R$. Thus it resembles coexistence, in that there is a stable mixed state; and coordination in that there are two stable fixed points.
\end{enumerate}
These regimes are all familiar in the evolutionary game theory literature for well-mixed populations. The first three correspond to the behaviour of two-player matrix games in well-mixed populations \cite{Traulsen2009}: (i) prisoner's dilemma/harmony game, (ii) snowdrift game, and (iii) stag-hunt game. The final type, coexistence \& coordination, arises for both the N-player stag-hunt \cite{Pacheco2009} and N-player snowdrift games \cite{Souza2009}.

For the well-mixed population we see dominance when $s$ is sufficiently small, and thus the PGG is approximating an NPD. As expected, cooperation is dominant when $b/c$ is sufficiently high. For higher values of $s$ there is a wide range of behaviour. In a region around $h=0.5$, if $b/c$ is large enough, there are coexistence \& coordination dynamics. There is a large basin of attraction for $n_R$  and if the system reaches this fixed point it will spend a long time in the vicinity. However, a single mutant invader must cross $n_L$ to reach this, against the selection pressure. As $b/c$ is increased, $n_L$ and $n_R$ move further apart, increasing the size of the basin of attraction for $n_R$. For $h=0.5$, the gradient of selection is symmetric ($n_L=Z-n_R$).

Decreasing $h$ from $0.5$, causes $n_L$ and $n_R$ to move to the left, eventually entering the coexistence regime. The basin of attraction for the internal stable fixed point is now $0<n<Z$. The system may spend a large amount of time near this point, although it will ultimately end up in one of the absorbing states. In the coexistence regime, as we discussed in \Cref{sec:epithelium_beneficial_coop} for the VD game with $h=0$, cooperators and defectors have a selective advantage when they are in sufficiently small numbers. This can lead to the case where both are beneficial mutants, and thus cooperation is able to be beneficial but not favoured.

Conversely as $h$ is increased from $0.5$, $n_L$ and $n_R$ move to the right and we enter the coordination regime. This corresponds to the region in \Cref{fig:coop_success} where very high values of the benefit-to-cost ratio are required for cooperation to be beneficial, even when cooperation is favoured. In \Cref{sec:epithelium_beneficial_coop} we argued, for the VD game with $h=0.5$, that this is due to the fact that both cooperators and defectors are at a disadvantage when in small numbers. Indeed this is the defining feature of the coordination regime, that $n=0$ and $n=1$ are stable fixed points. Thus any invader is at a disadvantage initially, as selection pushes it towards dying out. Therefore it is possible that defectors and cooperators can be at an evolutionary disadvantage compared to a neutral mutant.

The VT model behaviour is qualitatively very similar to that of the well-mixed population. The major difference is that the full spectrum of behaviour is available for a much smaller range of $b/c$ values for the VT model. This means that cooperation is successful at smaller benefit-to-cost ratios, as is consistent with our previous findings. It should be noted however, that these classifications are often approximate for the VT model. We observe, in a number of cases, additional fixed points very close to $n=0$ and $n=Z$. It is also clear from \Cref{fig:gos} that the coexistence \& coordination behaviour is much less pronounced that it is for the well-mixed case, with the internal fixed points much closer to the boundaries.

\section{Discussion} 
\label{sec:discussion}
There is an extensive literature on cancer modelling, which goes way beyond evolutionary game theory. For a review, see for example \mbox{\cite{Altrock2015}}. However, evolutionary game theory is increasingly used in cancer modelling \mbox{\cite{Rockne2019,Archetti2019,Wolfl2020}} both to elucidate tumorigenesis \mbox{\cite{Tomlinson1997,Bach2003,Basanta2008a,Basanta2008,Archetti2016}} and to inform potential treatment strategies \mbox{\cite{Basanta2012a,Kaznatcheev2017,Zhang,West2018}}. Experimental evidence that malignant cells cooperate to drive tumour growth has been found for breast cancer \mbox{\cite{Marusyk2014,Cleary2014}} and glioblastoma \mbox{\cite{Inda2010}}. Furthermore, evolutionary games have been explicitly quantified in non-small cell lung cancer \mbox{\cite{Kaznatcheev2019}} and neuroendocrine pancreatic cancer cell cultures \mbox{\cite{Archetti2015}}. These cancers both originate in epithelial cells, of the lung and pancreas, respectively. Disrupting cooperation could therefore be important for improving cancer treatment \mbox{\cite{Zhou2017}}.

Many models of cancer evolution assume populations of cells to be well-mixed  \mbox{\cite{Basanta2012a,Archetti2013b,Gerlee2017,West2018}}. However, the importance of spatial structure is increasingly recognised, even for simple mutations \mbox{\cite{Waclaw2015,West2021}}.
Population structure has long been established as a mechanism for promoting the evolution of cooperation \cite{Nowak2006}. If interactions are limited to an individual's neighbourhood, then cooperators can form mutually beneficial clusters. However, the success of cooperation is also influenced by the update rule. Results for the cycle graph in \Cref{sec:favoured_coop} demonstrate that a global update rule, such as the shift update, can lead to less stringent conditions for cooperation to be favoured when compared to local update rules. Within the local update rules there are also clear differences: cooperation tends to fare better with a death-birth update rule than a birth-death. In fact for the birth-death update on a cycle, the condition for cooperation to be favoured under an NPD game is equivalent to the well-mixed population. Thus the benefits of clustering are negated in this case.

It is therefore important to take into account realistic population structure and update dynamics of the tissue or tumour when considering the evolution of cooperation amongst cells. Our focus has been to consider how global updating affects the evolution of cooperation in a population structure representative of an epithelium. We have used the VT model to represent an epithelium, which allows death and division to be implemented independently, and therefore, it is trivial to implement what we call the decoupled update rule \mbox{\cite{Renton2019}}. We chose to focus on global updating, as it presents the opposite extreme to local update rules which have been extensively studied within evolutionary graph theory \mbox{\cite{Zukewich2013,Allen2016,Pena2016}}. Furthermore, we have been able to derive quasi-analytical results, which could be applied to other population models which use global updating. Our results are general for multiplayer games; however, we have focused on the analysis of sigmoid public goods games, as it has been proposed that they provide good models for the production of diffusible growth factors.

We have demonstrated that, for a sigmoid PGG, cooperation is more successful in the VT model compared to a well-mixed population. In both cases, the evolutionary outcomes depend on the parameters $s$ and $h$ of the logistic benefit function, as well as the benefit-to-cost ratio. In general, a lower benefit-to-cost ratio is required for cooperative success for the VT model, than the well-mixed population. In other words cells need a lower incentive to cooperate. This is consistent with our expectations: both models use global updating, however the population structure in the VT model allows for positive assortment of cooperators.

Although cooperation is more successful in the VT model, than the well-mixed population, the qualitative behaviour is very similar. We have characterised the evolutionary dynamics by considering conditions for which cooperation is beneficial and/or favourable, as well as calculating the gradient of selection.

As long as the steepness, $s$, is large enough, we tend to see coexistence behaviour when the inflection point, $h$, is less than a half and coordination behaviour when it is greater. These regimes are characterised by the fixed points of the gradient of selection. They also correspond to the regions in parameter space where cooperation is beneficial, but not favourable (coexistence), and favourable, but not beneficial (coordination). For small steepness, the game approaches an NPD and there is dominance behaviour. In this regime, conditions for cooperation to be beneficial and favoured coincide.

Examining the gradient of selection enables us to identify an additional dynamical regime: mixed coexistence \& coordination, which occurs in a region around $h=0.5$, as long as $s$ and $b/c$ are sufficiently large. This regime is characterised by two stable fixed points, one corresponding to all-defection, and the other to a heterogenous, majority-cooperator state. This dynamic has been identified previously for both well-mixed populations \cite{Archetti2013b} and graph-structured populations with local updating \cite{Archetti2016}. We have shown that it can also occur for the VT model, however the internal fixed points tend to be much closer to the boundaries.

It is beyond the scope of this paper to consider the full dynamics for an epithelial population structure with local update rules. However, we have considered conditions for cooperation to be favourable on a hexagonal lattice with death-birth update rule using results from \cite{Pena2016}. We found the critical benefit-to-cost ratios to be intermediate between the well-mixed population and VT model. This is consistent with previous results for two-player games \cite{Renton2019}. Taken together, these results suggest not only that population structure promotes cooperation, but that global updating also plays a crucial role. We can thus consider a general rule for cooperation is that it prefers local game play but global competition for offspring.

It is worth taking a moment to consider the implications of beneficial and favourable mutations for invasion, and how we distinguish between the two concepts. Whether or not a mutation is beneficial is perhaps the most relevant measure for a single invasion event. It essentially tells us that the mutated cell has a higher probability of invasion in a wild-type population than a wild-type cell would have, and therefore it has an evolutionary advantage. The significance of a mutation being favourable is a little less clear, as it compares two different invasion processes: the probability of invasion of a mutated cell in a wild-type population is higher than the converse scenario, where a wild-type cell invades a population of mutants. However, the condition for a mutant to be favoured is also equivalent to the condition for cooperation to dominate, if mutation is allowed.

These results suggest that the population structure, the update rule and the game all play important roles in determining the evolutionary success of cooperation. Cancer models which utilise evolutionary games \mbox{\cite{You2017,Gatenby2020}} may therefore underestimate the success of cooperative phenotypes, if they fail to account for population structure or assume that death and division are more tightly coupled than is realistic. For example, therapeutic strategies that aim to eliminate cooperation by manipulating evolutionary dynamics, rely on accurate predictions of those dynamics \mbox{\cite{Archetti2013}}. 

We do not suggest that our regime, with fully local interactions and fully global competition is necessarily realistic for invasion processes in tissues. We have assumed that cells only interact with their immediate neighbours; however, interaction groups may be much larger and likely depend on the specific context. For example, in the case of growth factor production, group size will depend on diffusion range. Estimates of these diffusion ranges are difficult to obtain experimentally \mbox{\cite{Archetti2020}}. However, larger group sizes tend to suppress cooperation \mbox{\cite{Archetti2012}}, so it is an important consideration.

We chose to focus on global updating, as it presents the opposite extreme to a local update rule. It is likely, however, that update dynamics in a real epithelium lie somewhere in between. Contact inhibition \mbox{\cite{McClatchey2012}}, and other density-dependent processes \mbox{\cite{Eisenhoffer2012,Fernandez-Gonzalez2012}}, result in spatial coupling between death and division, which is likely tissue dependent. Stronger contact inhibition could result in dynamics closer to the death-birth update \mbox{\cite{Mesa2018}}, while weaker contact inhibition is closer to global updating.

Interestingly, loss of contact inhibition is associated with malignancy \mbox{\cite{McClatchey2012}}, suggesting that updating is more global, and thus cooperation could be more successful, than in healthy tissues. In future work, we will consider the effect of contact inhibition on cooperation, and the spectrum of behaviour between local and global updating. Understanding the nature of spatial coupling in real epithelia, or in cancerous tumours, could be crucial for predicting evolutionary outcomes.

Our general conclusion that local game play and global competition for offspring favour cooperation has implications beyond applications to cancer, where cooperation unusually may be considered undesirable. In a societal context, where cooperation is desirable, it may be promoted by engineering an environment rich in local social interactions, which nevertheless allows for imitation of successful strategies more globally.  

\vspace{10pt}
\begin{small}

\noindent 
\textbf{Data accessibility.} The code and data can be accessed at https://github.com/\linebreak jessierenton/pgg-epithelium \\
\noindent 
\textbf{Authors' contributions.} JR and KMP designed the research. JR carried out the research and wrote the paper. KMP edited the paper.  \\
\noindent 
\textbf{Acknowledgments.} This research was funded by an EPSRC studentship held by JR. \\
\noindent 
\textbf{Competing interests.} We declare we have no competing interests. \\

\end{small}

\bibliographystyle{num} 
\bibliography{bib}

\begin{thebibliography}{10}
\providecommand{\url}[1]{\texttt{#1}}
\providecommand{\urlprefix}{URL }
\providecommand{\eprint}[2][]{#2}

\bibitem{Hanahan2000}
D.~Hanahan and R.~A. Weinberg. {The hallmarks of cancer}. \emph{Cell},
  100:57--70, 2000, \url{http://dx.doi.org/10.1007/s00262-010-0968-0}.

\bibitem{Hanahan2011}
D.~Hanahan and R.~A. Weinberg. {Hallmarks of cancer: The next generation}.
  \emph{Cell}, 144(5):646--674, 2011,
  \url{http://dx.doi.org/10.1016/j.cell.2011.02.013}.

\bibitem{Jouanneau1994}
J.~Jouanneau, G.~Moens, Y.~Bourgeois, M.~F. Poupon, and J.~P. Thiery. {A
  minority of carcinoma cells producing acidic fibroblast growth factor induces
  a community effect for tumor progression}. \emph{Proceedings of the National
  Academy of Sciences of the United States of America}, 91(1):286--290, 1994,
  \url{http://dx.doi.org/10.1073/pnas.91.1.286}.

\bibitem{Axelrod2006}
R.~Axelrod, D.~E. Axelrod, and K.~J. Pienta. {Evolution of cooperation among
  tumor cells}. \emph{Proceedings of the National Academy of Sciences of the
  United States of America}, 103(36):13474--13479, 2006,
  \url{http://dx.doi.org/10.1073/pnas.0606053103}.

\bibitem{Warburg1956}
O.~Warburg. {On the origin of cancer cells}. \emph{Science},
  123(3191):309--314, 1956,
  \url{http://dx.doi.org/10.1126/science.123.3191.309}.

\bibitem{Archetti2014}
M.~Archetti. {Evolutionary dynamics of the Warburg effect: Glycolysis as a
  collective action problem among cancer cells}. \emph{Journal of Theoretical
  Biology}, 341:1--8, 2014, \url{http://dx.doi.org/10.1016/j.jtbi.2013.09.017}.

\bibitem{Nowak2006}
M.~A. Nowak. {Five rules for the evolution of cooperation}. \emph{Science},
  314(5805):1560--1563, 2006, \url{http://dx.doi.org/10.1126/science.1133755}.

\bibitem{Ohtsuki2006a}
H.~Ohtsuki, C.~Hauert, E.~Lieberman, and M.~A. Nowak. {A simple rule for the
  evolution of cooperation on graphs and social networks}. \emph{Nature},
  441(7092):502--505, 2006, \url{http://dx.doi.org/10.1038/nature04605}.

\bibitem{Allen2016}
B.~Allen, G.~Lippner, Y.~T. Chen, B.~Fotouhi, N.~Momeni, S.~T. Yau, and M.~A.
  Nowak. {Evolutionary dynamics on any population structure}. \emph{Nature},
  544(7649):227--230, 2017, \url{http://dx.doi.org/10.1038/nature21723}.

\bibitem{Marusyk2014}
A.~Marusyk, D.~P. Tabassum, P.~M. Altrock, V.~Almendro, F.~Michor, and
  K.~Polyak. {Non-cell-autonomous driving of tumour growth supports sub-clonal
  heterogeneity}. \emph{Nature}, 2014,
  \url{http://dx.doi.org/10.1038/nature13556}.

\bibitem{Archetti2013}
M.~Archetti. {Evolutionarily stable anti-cancer therapies by autologous cell
  defection}. \emph{Evolution, Medicine, and Public Health}, 2013(1):161--172,
  2013, \url{http://dx.doi.org/10.1093/emph/eot014}.

\bibitem{Zhou2017}
H.~Zhou, D.~Neelakantan, and H.~L. Ford. {Clonal cooperativity in heterogenous
  cancers}. \emph{Seminars in Cell and Developmental Biology}, 64:79--89, 2017,
  \url{http://dx.doi.org/10.1016/j.semcdb.2016.08.028}.

\bibitem{Tomlinson1997b}
I.~P.~M. Tomlinson. {Game-theory models of interactions between tumour cells}.
  \emph{European Journal of Cancer Part A}, 33(9):1495--1500, 1997,
  \url{http://dx.doi.org/10.1016/S0959-8049(97)00170-6}.

\bibitem{Basanta}
D.~Basanta and A.~Deutsch. {A game theoretical perspective on the somatic
  evolution of cancer}. In \emph{Selected topics in cancer modeling: Genesis,
  evolution, immune competition, and therapy}, 1--16, Birkh{\"{a}}user Boston,
  2008, \eprint{0810.4738},
  \url{http://dx.doi.org/10.1007/978-0-8176-4713-1_5}.

\bibitem{Hummert2014}
S.~Hummert, K.~Bohl, D.~Basanta, A.~Deutsch, S.~Werner, G.~Thei{\ss}en,
  A.~Schroeter, and S.~Schuster. {Evolutionary game theory: Cells as players}.
  \emph{Molecular BioSystems}, 10(12):3044--3065, 2014,
  \url{http://dx.doi.org/10.1039/c3mb70602h}.

\bibitem{Archetti2012}
M.~Archetti and I.~Scheuring. {Review: Game theory of public goods in one-shot
  social dilemmas without assortment}. \emph{Journal of Theoretical Biology},
  299:9--20, 2012, \url{http://dx.doi.org/10.1016/j.jtbi.2011.06.018}.

\bibitem{Hauert2002}
C.~Hauert, S.~{De Monte}, J.~Hofbauer, and K.~Sigmund. {Volunteering as Red
  Queen mechanism for cooperation in public goods games}. \emph{Science},
  296(5570):1129--1132, 2002, \url{http://dx.doi.org/10.1126/science.1070582}.

\bibitem{Santos2008}
F.~C. Santos, M.~D. Santos, and J.~M. Pacheco. {Social diversity promotes the
  emergence of cooperation in public goods games}. \emph{Nature},
  454(7201):213--216, 2008, \url{http://dx.doi.org/10.1038/nature06940}.

\bibitem{Archetti2015}
M.~Archetti, D.~A. Ferraro, and G.~Christofori. {Heterogeneity for IGF-II
  production maintained by public goods dynamics in neuroendocrine pancreatic
  cancer}. \emph{Proceedings of the National Academy of Sciences of the United
  States of America}, 112(6):1833--1838, 2015,
  \url{http://dx.doi.org/10.1073/pnas.1414653112}.

\bibitem{Archetti2020}
M.~Archetti, I.~Scheuring, and D.~W. Yu. {The non-tragedy of the non-linear
  commons}. \emph{Preprints}, \eprint{2020040226}, 2020,
  \url{http://dx.doi.org/10.20944/preprints202004.0226.v1}.

\bibitem{Bach2001}
L.~A. Bach, S.~M. Bentzen, J.~Alsner, and F.~B. Christiansen. {An
  evolutionary-game model of tumour-cell interactions: Possible relevance to
  gene therapy}. \emph{European Journal of Cancer}, 37(16):2116--2120, 2001,
  \url{http://dx.doi.org/10.1016/S0959-8049(01)00246-5}.

\bibitem{Bach2006}
L.~A. Bach, T.~Helvik, and F.~B. Christiansen. {The evolution of n-player
  cooperation - Threshold games and ESS bifurcations}. \emph{Journal of
  Theoretical Biology}, 238(2):426--434, 2006,
  \url{http://dx.doi.org/10.1016/j.jtbi.2005.06.007}.

\bibitem{Archetti2009a}
M.~Archetti. {The volunteer's dilemma and the optimal size of a social group}.
  \emph{Journal of Theoretical Biology}, 261(3):475--480, 2009,
  \url{http://dx.doi.org/10.1016/j.jtbi.2009.08.018}.

\bibitem{Archetti2009}
M.~Archetti. {Cooperation as a volunteer's dilemma and the strategy of conflict
  in public goods games}. \emph{Journal of Evolutionary Biology},
  22(11):2192--2200, 2009,
  \url{http://dx.doi.org/10.1111/j.1420-9101.2009.01835.x}.

\bibitem{Archetti2011}
M.~Archetti and I.~Scheuring. {Coexistence of cooperation and defection in
  public goods games}. \emph{Evolution}, 65(4):1140--1148, 2011,
  \url{http://dx.doi.org/10.1111/j.1558-5646.2010.01185.x}.

\bibitem{Ohtsuki2006}
H.~Ohtsuki, C.~Hauert, E.~Lieberman, and M.~A. Nowak. {A simple rule for the
  evolution of cooperation on graphs and social networks}. \emph{Nature},
  441(7092):502--505, 2006, \url{http://dx.doi.org/10.1038/nature04605}.

\bibitem{Nowak2010}
M.~A. Nowak, C.~E. Tarnita, and T.~Antal. {Evolutionary dynamics in structured
  populations}. \emph{Philosophical Transactions of the Royal Society B:
  Biological Sciences}, 365(1537):19--30, 2010,
  \url{http://dx.doi.org/10.1098/rstb.2009.0215}.

\bibitem{Pena2016}
J.~Pe{\~{n}}a, B.~Wu, J.~Arranz, and A.~Traulsen. {Evolutionary Games of
  Multiplayer Cooperation on Graphs}. \emph{PLoS Computational Biology},
  12(8):1--15, 2016, \url{http://dx.doi.org/10.1371/journal.pcbi.1005059}.

\bibitem{Lieberman2005}
E.~Lieberman, C.~Hauert, and M.~A. Nowak. {Evolutionary dynamics on graphs}.
  \emph{Nature 2004 433:7023}, 433(7023):312--316, 2005,
  \url{http://dx.doi.org/10.1038/nature03204}.

\bibitem{Archetti2016}
M.~Archetti. {Cooperation among cancer cells as public goods games on Voronoi
  networks}. \emph{Journal of Theoretical Biology}, 396:191--203, 2016,
  \url{http://dx.doi.org/10.1016/j.jtbi.2016.02.027}.

\bibitem{Renton2019}
J.~Renton and K.~M. Page. {Evolution of cooperation in an epithelium}.
  \emph{Journal of the Royal Society Interface}, 16(152):20180918, 2019,
  \url{http://dx.doi.org/10.1098/rsif.2018.0918}.

\bibitem{Nathanson2009}
C.~G. Nathanson, C.~E. Tarnita, and M.~A. Nowak. {Calculating evolutionary
  dynamics in structured populations}. \emph{PLoS Computational Biology},
  5(12):e1000615, 2009, \url{http://dx.doi.org/10.1371/journal.pcbi.1000615}.

\bibitem{Zukewich2013}
J.~Zukewich, V.~Kurella, M.~Doebeli, and C.~Hauert. {Consolidating birth-death
  and death-birth processes in structured populations}. \emph{PLoS ONE},
  8(1):e54639, 2013, \url{http://dx.doi.org/10.1371/journal.pone.0054639}.

\bibitem{Masuda2009}
N.~Masuda. {Directionality of contact networks suppresses selection pressure in
  evolutionary dynamics}. \emph{Journal of Theoretical Biology},
  258(2):323--334, 2009, \url{http://dx.doi.org/10.1016/j.jtbi.2009.01.025}.

\bibitem{Antal2009}
T.~Antal, H.~Ohtsuki, J.~Wakeley, P.~D. Taylor, and M.~A. Nowak. {Evolution of
  cooperation by phenotypic similarity}. \emph{Proceedings of the National
  Academy of Sciences of the United States of America}, 106(21):8597--8600,
  2009, \url{http://dx.doi.org/10.1073/pnas.0902528106}.

\bibitem{Tarnita2009a}
C.~E. Tarnita, T.~Antal, H.~Ohtsuki, and M.~A. Nowak. {Evolutionary dynamics in
  set structured populations}. \emph{Proceedings of the National Academy of
  Sciences of the United States of America}, 106(21):8601--8604, 2009,
  \url{http://dx.doi.org/10.1073/pnas.0903019106}.

\bibitem{Allen2012}
B.~Allen and M.~A. Nowak. {Evolutionary shift dynamics on a cycle}.
  \emph{Journal of Theoretical Biology}, 311:28--39, 2012,
  \url{http://dx.doi.org/10.1016/j.jtbi.2012.07.006}.

\bibitem{Pavlogiannis2015}
A.~Pavlogiannis, K.~Chatterjee, B.~Adlam, and M.~A. Nowak. {Cellular
  cooperation with shift updating and repulsion}. \emph{Scientific Reports},
  5(17147), 2015, \url{http://dx.doi.org/10.1038/srep17147}.

\bibitem{Mesa2018}
K.~R. Mesa, K.~Kawaguchi, K.~Cockburn, D.~Gonzalez, J.~Boucher, T.~Xin, A.~M.
  Klein, and V.~Greco. {Homeostatic epidermal stem cell self-renewal is driven
  by local differentiation}. \emph{Cell Stem Cell}, 23(5):677--686, 2018,
  \url{http://dx.doi.org/10.1016/j.stem.2018.09.005}.

\bibitem{Maciejewski2014a}
W.~Maciejewski, F.~Fu, and C.~Hauert. {Evolutionary Game Dynamics in
  Populations with Heterogenous Structures}. \emph{PLOS Computational Biology},
  10(4):e1003567, 2014, \url{http://dx.doi.org/10.1371/JOURNAL.PCBI.1003567}.

\bibitem{Meineke2001}
F.~A. Meineke, C.~S. Potten, and M.~Loeffler. {Cell migration and organization
  in the intestinal crypt using a lattice-free model}. \emph{Cell
  Proliferation}, 34(4):253--266, 2001,
  \url{http://dx.doi.org/10.1046/j.0960-7722.2001.00216.x}.

\bibitem{VanLeeuwen2009}
I.~M.~M. {Van Leeuwen}, G.~R. Mirams, A.~Walter, A.~Fletcher, P.~Murray,
  J.~Osborne, S.~Varma, S.~J. Young, J.~Cooper, B.~Doyle, J.~Pitt-Francis,
  L.~Momtahan, P.~Pathmanathan, J.~P. Whiteley, S.~J. Chapman, D.~J. Gavaghan,
  O.~E. Jensen, J.~R. King, P.~K. Maini, S.~L. Waters, and H.~M. Byrne. {An
  integrative computational model for intestinal tissue renewal}. \emph{Cell
  Proliferation}, 42(5):617--636, 2009,
  \url{http://dx.doi.org/10.1111/j.1365-2184.2009.00627.x}.

\bibitem{Mirams2012}
G.~R. Mirams, A.~G. Fletcher, P.~K. Maini, and H.~M. Byrne. {A theoretical
  investigation of the effect of proliferation and adhesion on monoclonal
  conversion in the colonic crypt}. \emph{Journal of Theoretical Biology},
  312:143--156, 2012, \url{http://dx.doi.org/10.1016/j.jtbi.2012.08.002}.

\bibitem{Romijn}
L.~B. Romijn, A.~A. Almet, C.~W. Tan, and J.~M. Osborne. {Modelling the effect
  of subcellular mutations on the migration of cells in the colorectal crypt}.
  \emph{BMC Bioinformatics}, 21(1), 2020,
  \url{http://dx.doi.org/10.1186/s12859-020-3391-3}.

\bibitem{Farhadifar2007}
R.~Farhadifar, J.~C. R{\"{o}}per, B.~Aigouy, S.~Eaton, and F.~J{\"{u}}licher.
  {The influence of cell mechanics, cell-cell interactions, and proliferation
  on epithelial packing}. \emph{Current Biology}, 17(24):2095--2104, 2007,
  \url{http://dx.doi.org/10.1016/j.cub.2007.11.049}.

\bibitem{Curtius2017}
K.~Curtius, N.~A. Wright, and T.~A. Graham. {An evolutionary perspective on
  field cancerization}. \emph{Nature Reviews Cancer}, 18(1):19--32, 2017,
  \url{http://dx.doi.org/10.1038/nrc.2017.102}.

\bibitem{Moran1958}
P.~A. Moran. {Random processes in genetics}. \emph{Mathematical Proceedings of
  the Cambridge Philosophical Society}, 54(1):60--71, 1958,
  \url{http://dx.doi.org/10.1017/S0305004100033193}.

\bibitem{Wu2013}
B.~Wu, A.~Traulsen, and C.~S. Gokhale. {Dynamic properties of evolutionary
  multi-player games in finite populations}. \emph{Games}, 4(2):182--199, 2013,
  \url{http://dx.doi.org/10.3390/g4020182}.

\bibitem{Traulsen2009}
A.~Traulsen and C.~Hauert. {Stochastic Evolutionary Game Dynamics}. In H.~G.
  Schuster, ed., \emph{Reviews of nonlinear dynamics and complexity}, vol.~2,
  chap.~1, 25--61, Wiley-VCH, 2009,
  \url{http://dx.doi.org/10.1016/B978-0-444-53766-9.00006-9}.

\bibitem{Tarnita2009}
C.~E. Tarnita, H.~Ohtsuki, T.~Antal, F.~Fu, and M.~A. Nowak. {Strategy
  selection in structured populations}. \emph{Journal of Theoretical Biology},
  259(3):570--581, 2009, \url{http://dx.doi.org/10.1016/j.jtbi.2009.03.035}.

\bibitem{Gokhale2010}
C.~S. Gokhale and A.~Traulsen. {Evolutionary games in the multiverse}.
  \emph{Proceedings of the National Academy of Sciences of the United States of
  America}, 107(12):5500--5504, 2010,
  \url{http://dx.doi.org/10.1073/pnas.0912214107}.

\bibitem{Archetti2013b}
M.~Archetti. {Evolutionary game theory of growth factor production:
  Implications for tumour heterogeneity and resistance to therapies}.
  \emph{British Journal of Cancer}, 109(4):1056--1062, 2013,
  \url{http://dx.doi.org/10.1038/bjc.2013.336}.

\bibitem{Archetti2013a}
M.~Archetti. {Dynamics of growth factor production in monolayers of cancer
  cells and evolution of resistance to anticancer therapies}.
  \emph{Evolutionary Applications}, 6(8):1146--1159, 2013,
  \url{http://dx.doi.org/10.1111/eva.12092}.

\bibitem{Pacheco2009}
J.~M. Pacheco, F.~C. Santos, M.~O. Souza, and B.~Skyrms. {Evolutionary dynamics
  of collective action in N-person stag hunt dilemmas}. \emph{Proceedings of
  the Royal Society B: Biological Sciences}, 276(1655):315--321, 2009,
  \url{http://dx.doi.org/10.1098/rspb.2008.1126}.

\bibitem{Souza2009}
M.~O. Souza, J.~M. Pacheco, and F.~C. Santos. {Evolution of cooperation under
  N-person snowdrift games}. \emph{Journal of Theoretical Biology},
  260(4):581--588, 2009, \url{http://dx.doi.org/10.1016/j.jtbi.2009.07.010}.

\bibitem{Altrock2015}
P.~M. Altrock, L.~L. Liu, and F.~Michor. {The mathematics of cancer:
  Integrating quantitative models}. \emph{Nature Reviews Cancer},
  15(12):730--745, 2015, \url{http://dx.doi.org/10.1038/nrc4029}.

\bibitem{Rockne2019}
R.~C. Rockne, A.~Hawkins-Daarud, K.~R. Swanson, J.~P. Sluka, J.~A. Glazier,
  P.~Macklin, D.~A. Hormuth, A.~M. Jarrett, E.~A. Lima, J.~{Tinsley Oden},
  G.~Biros, T.~E. Yankeelov, K.~Curtius, I.~{Al Bakir}, D.~Wodarz, N.~Komarova,
  L.~Aparicio, M.~Bordyuh, R.~Rabadan, S.~D. Finley, H.~Enderling, J.~Caudell,
  E.~G. Moros, A.~R. Anderson, R.~A. Gatenby, A.~Kaznatcheev, P.~Jeavons,
  N.~Krishnan, J.~Pelesko, R.~R. Wadhwa, N.~Yoon, D.~Nichol, A.~Marusyk,
  M.~Hinczewski, and J.~G. Scott. {The 2019 mathematical oncology roadmap}.
  \emph{Physical Biology}, 16(4):41005, 2019,
  \url{http://dx.doi.org/10.1088/1478-3975/ab1a09}.

\bibitem{Archetti2019}
M.~Archetti and K.~J. Pienta. {Cooperation among cancer cells: applying game
  theory to cancer}. \emph{Nature Reviews Cancer}, 19(2):110--117, 2019,
  \url{http://dx.doi.org/10.1038/s41568-018-0083-7}.

\bibitem{Wolfl2020}
B.~W{\"{o}}lfl, H.~te~Rietmole, M.~Salvioli, F.~Thuijsman, J.~S. Brown,
  B.~Burgering, and K.~Staňkov{\'{a}}. {The contribution of evolutionary game
  theory to understanding and treating cancer}. \emph{medRxiv},
  \eprint{2020.12.02.20241703}, 2020,
  \url{http://dx.doi.org/10.1101/2020.12.02.20241703}.

\bibitem{Tomlinson1997}
I.~P.~M. Tomlinson and W.~F. Bodmer. {Modelling the consequences of
  interactions between tumour cells}. \emph{British Journal of Cancer},
  75(2):157--160, 1997, \url{http://dx.doi.org/10.1038/bjc.1997.26}.

\bibitem{Bach2003}
L.~A. Bach, D.~J. Sumpter, J.~Alsner, and V.~Loeschcke. {Spatial evolutionary
  games of interaction among generic cancer cells}. \emph{Journal of
  Theoretical Medicine}, 5(1):47--58, 2003,
  \url{http://dx.doi.org/10.1080/10273660310001630443}.

\bibitem{Basanta2008a}
D.~Basanta, H.~Hatzikirou, and A.~Deutsch. {Studying the emergence of
  invasiveness in tumours using game theory}. \emph{European Physical Journal
  B}, 63(3):393--397, 2008, \url{http://dx.doi.org/10.1140/epjb/e2008-00249-y}.

\bibitem{Basanta2008}
D.~Basanta, M.~Simon, H.~Hatzikirou, and A.~Deutsch. {Evolutionary game theory
  elucidates the role of glycolysis in glioma progression and invasion}.
  \emph{Cell Proliferation}, 41(6):980--987, 2008,
  \url{http://dx.doi.org/10.1111/j.1365-2184.2008.00563.x}.

\bibitem{Basanta2012a}
D.~Basanta, R.~A. Gatenby, and A.~R. Anderson. {Exploiting evolution to treat
  drug resistance: Combination therapy and the double bind}. \emph{Molecular
  Pharmaceutics}, 9(4):914--921, 2012,
  \url{http://dx.doi.org/10.1021/mp200458e}.

\bibitem{Kaznatcheev2017}
A.~Kaznatcheev, R.~{Vander Velde}, J.~G. Scott, and D.~Basanta. {Cancer
  treatment scheduling and dynamic heterogeneity in social dilemmas of tumour
  acidity and vasculature}. \emph{British Journal of Cancer}, 116(6):785--792,
  2017, \url{http://dx.doi.org/10.1038/bjc.2017.5}.

\bibitem{Zhang}
J.~Zhang, J.~J. Cunningham, J.~S. Brown, and R.~A. Gatenby. {Integrating
  evolutionary dynamics into treatment of metastatic castrate-resistant
  prostate cancer}. \emph{Nature Communications}, 8(1), 2017,
  \url{http://dx.doi.org/10.1038/s41467-017-01968-5}.

\bibitem{West2018}
J.~West, Y.~Ma, and P.~K. Newton. {Capitalizing on competition: An evolutionary
  model of competitive release in metastatic castration resistant prostate
  cancer treatment}. \emph{Journal of Theoretical Biology}, 455:249--260, 2018,
  \url{http://dx.doi.org/10.1016/j.jtbi.2018.07.028}.

\bibitem{Cleary2014}
A.~S. Cleary, T.~L. Leonard, S.~A. Gestl, and E.~J. Gunther. {Tumour cell
  heterogeneity maintained by cooperating subclones in Wnt-driven mammary
  cancers}. \emph{Nature}, 508(1):113--117, 2014,
  \url{http://dx.doi.org/10.1038/nature13187}.

\bibitem{Inda2010}
M.~D.~M. Inda, R.~Bonavia, A.~Mukasa, Y.~Narita, D.~W. Sah, S.~Vandenberg,
  C.~Brennan, T.~G. Johns, R.~Bachoo, P.~Hadwiger, P.~Tan, R.~A. DePinho,
  W.~Cavenee, and F.~Furnari. {Tumor heterogeneity is an active process
  maintained by a mutant EGFR-induced cytokine circuit in glioblastoma}.
  \emph{Genes and Development}, 24(16):1731--1745, 2010,
  \url{http://dx.doi.org/10.1101/gad.1890510}.

\bibitem{Kaznatcheev2019}
A.~Kaznatcheev, J.~Peacock, D.~Basanta, A.~Marusyk, and J.~G. Scott.
  {Fibroblasts and alectinib switch the evolutionary games played by non-small
  cell lung cancer}. \emph{Nature Ecology and Evolution}, 3(3):450--456, 2019,
  \url{http://dx.doi.org/10.1038/s41559-018-0768-z}.

\bibitem{Gerlee2017}
P.~Gerlee and P.~M. Altrock. {Extinction rates in tumour public goods games}.
  \emph{Journal of the Royal Society Interface}, 14(134), 2017,
  \url{http://dx.doi.org/10.1098/rsif.2017.0342}.

\bibitem{Waclaw2015}
B.~Waclaw, I.~Bozic, M.~E. Pittman, R.~H. Hruban, B.~Vogelstein, and M.~A.
  Nowak. {A spatial model predicts that dispersal and cell turnover limit
  intratumour heterogeneity}. \emph{Nature}, 525(7568):261--264, 2015,
  \url{http://dx.doi.org/10.1038/nature14971}.

\bibitem{West2021}
J.~West, R.~O. Schenck, C.~Gatenbee, M.~Robertson-Tessi, and A.~R.~A. Anderson.
  {Normal tissue architecture determines the evolutionary course of cancer}.
  \emph{Nature Communications}, 12(1):1--9, 2021,
  \url{http://dx.doi.org/10.1038/s41467-021-22123-1}.

\bibitem{You2017}
L.~You, J.~S. Brown, F.~Thuijsman, J.~J. Cunningham, R.~A. Gatenby, J.~Zhang,
  and K.~Staňkov{\'{a}}. {Spatial vs. non-spatial eco-evolutionary dynamics in
  a tumor growth model}. \emph{Journal of Theoretical Biology}, 435:78--97,
  2017, \url{http://dx.doi.org/10.1016/j.jtbi.2017.08.022}.

\bibitem{Gatenby2020}
R.~A. Gatenby and J.~S. Brown. {Integrating evolutionary dynamics into cancer
  therapy}. \emph{Nature Reviews Clinical Oncology}, 17(11):675--686, 2020,
  \url{http://dx.doi.org/10.1038/s41571-020-0411-1}.

\bibitem{McClatchey2012}
A.~I. McClatchey and A.~S. Yap. {Contact inhibition (of proliferation) redux}.
  \emph{Current Opinion in Cell Biology}, 24(5):685--694, 2012,
  \url{http://dx.doi.org/10.1016/j.ceb.2012.06.009}.

\bibitem{Eisenhoffer2012}
G.~T. Eisenhoffer, P.~D. Loftus, M.~Yoshigi, H.~Otsuna, C.~B. Chien, P.~A.
  Morcos, and J.~Rosenblatt. {Crowding induces live cell extrusion to maintain
  homeostatic cell numbers in epithelia}. \emph{Nature}, 484(7395):546--549,
  2012, \url{http://dx.doi.org/10.1038/nature10999}.

\bibitem{Fernandez-Gonzalez2012}
R.~Fernandez-Gonzalez and J.~A. Zallen. {Feeling the squeeze: live-cell
  extrusion limits cell density in epithelia}. \emph{Cell}, 149(5):965--967,
  2012, \url{http://dx.doi.org/10.1016/j.cell.2012.05.006}.

\bibitem{Osborne2017}
J.~M. Osborne, A.~G. Fletcher, J.~M. Pitt-Francis, P.~K. Maini, and D.~J.
  Gavaghan. {Comparing individual-based approaches to modelling the
  self-organization of multicellular tissues}. \emph{PLoS Computational
  Biology}, 13(2):e1005387, 2017,
  \url{http://dx.doi.org/10.1371/journal.pcbi.1005387}.

\end{thebibliography}

\begin{appendices}
    \section{Voronoi tessellation model}
    \label{sec:VT_appendix}
    The Voronoi tessellation (VT) model was introduced in \cite{Meineke2001,VanLeeuwen2009}. We use the version and parameter values from \cite{Renton2019} in all simulations in this paper. Parameters are given in \Cref{table:params} and the model is defined as follows.
    \begin{table}[htb!]
        \caption{Table of parameters used in the Voronoi tessellation model \cite{Osborne2017,Renton2019}.}
        \label{table:params}
        \centering
        \begin{tabular}{lll}
            \toprule
            \textbf{Parameter} & \textbf{Description} & \textbf{Value}\\
            \midrule
                         $\mu$ &             spring constant &                           50\\
                         $s$   &             natural cell separation &                   1\\
                         $\epsilon$ &        initial sibling cell separation &           0.1\\
                         $\eta$ &            drag coefficient        &                   1\\
                         $\Delta t$ &        time-step (hours)           &               0.005\\
                         $\lambda$ &         rate of division/death (hours$^{-1}$)  &    $12^{-1}$\\
            \bottomrule
        \end{tabular}
    \end{table}
    
     The VT model represents a tissue as a set of points points in a fixed domain with periodic boundary conditions. Each point corresponds to a cell-centre and moves subject sto spring-like forces cells exert on their neighbours.
    
    We define the force cell $j$ exerts on its neighbour $i$ to be
    \begin{equation}
        \vec{F}_{ij}(t) = -\mu\vec{\hat{r}}_{ij}(t)(\abs{\vec{r}_{ij}(t)}-s_{ij}(t)) \, ,
    \end{equation}
    where $\mu$ is the spring constant, $\vec{r}_{ij}=\vec{r}_{i}-\vec{r}_j$ is the vector pointing from cell $j$ to cell $i$, and $s_{ij}$ is the natural separation between cells $i$ and $j$. This is set be a constant $s_{ij}=s$, with the exception that newborn sibling cells have a separation $\epsilon$ immediately after division. For these cells, $s_{ij}$ grows linearly over the course of an hour to reach $s$. The total force on a cell $i$ is given by
    \begin{equation}
        \vec{F}_i(t) = \sum_{j\in\mathcal{N}_i(t)} \vec{F}_{ij} \, ,
    \end{equation}
where $\mathcal{N}_i(t)$ denotes the set of cells neighbouring $i$.

    It is assumed that motion is over-damped and thus the equation of motion for $i$ is the first order differential equation
    \begin{equation}
        \eta \frac{d\vec{r}_i}{dt} = \vec{F}_i(t) \, ,
    \end{equation}
where $\eta$ is the damping constant. This is solved numerically using
    \begin{equation}
        \vec{r}_i(t+\Delta t) = \vec{r}_i(t)+\frac{\Delta t}{\eta}\vec{F}_i \, ,
    \end{equation}
where the time-step, $\Delta t$, is sufficiently small to ensure numerical stability.

At each time-step, after cells have moved, a Voronoi tessellation is performed. This partitions the domain into polygonal regions, each corresponding to the shape of a cell. It also defines the neighbourhood connections, which are needed to determine the forces cells exert on one another, as well as cell fitnesses. 

Cell division is implemented within the VT model, by removing the parent cell and replacing it with two progeny cells, separated by a distance $\epsilon$, across a uniformly random axis. Cell death simply requires the dead cell to be removed. After a death or division, Voronoi tessellation must be performed to obtain new neighbour connections.

\section{Neighbour distributions in the VT model}
\label{sec:neighbour_dist_appendix}
In \Cref{sec:favoured_coop,sec:beneficial_coop} we derived conditions under which cooperation is favoured and beneficial, given by \Cref{eq:global_coop_favoured,eq:global_coop_beneficial} respectively. These derivations are based on the assumption that the frequency of cells with $k$ neighbours is a fixed distribution, $g_k$, independent of the cell type or the number of cooperators in the population, $n$.

\begin{figure}[htb!]
    \includegraphics[scale=0.7]{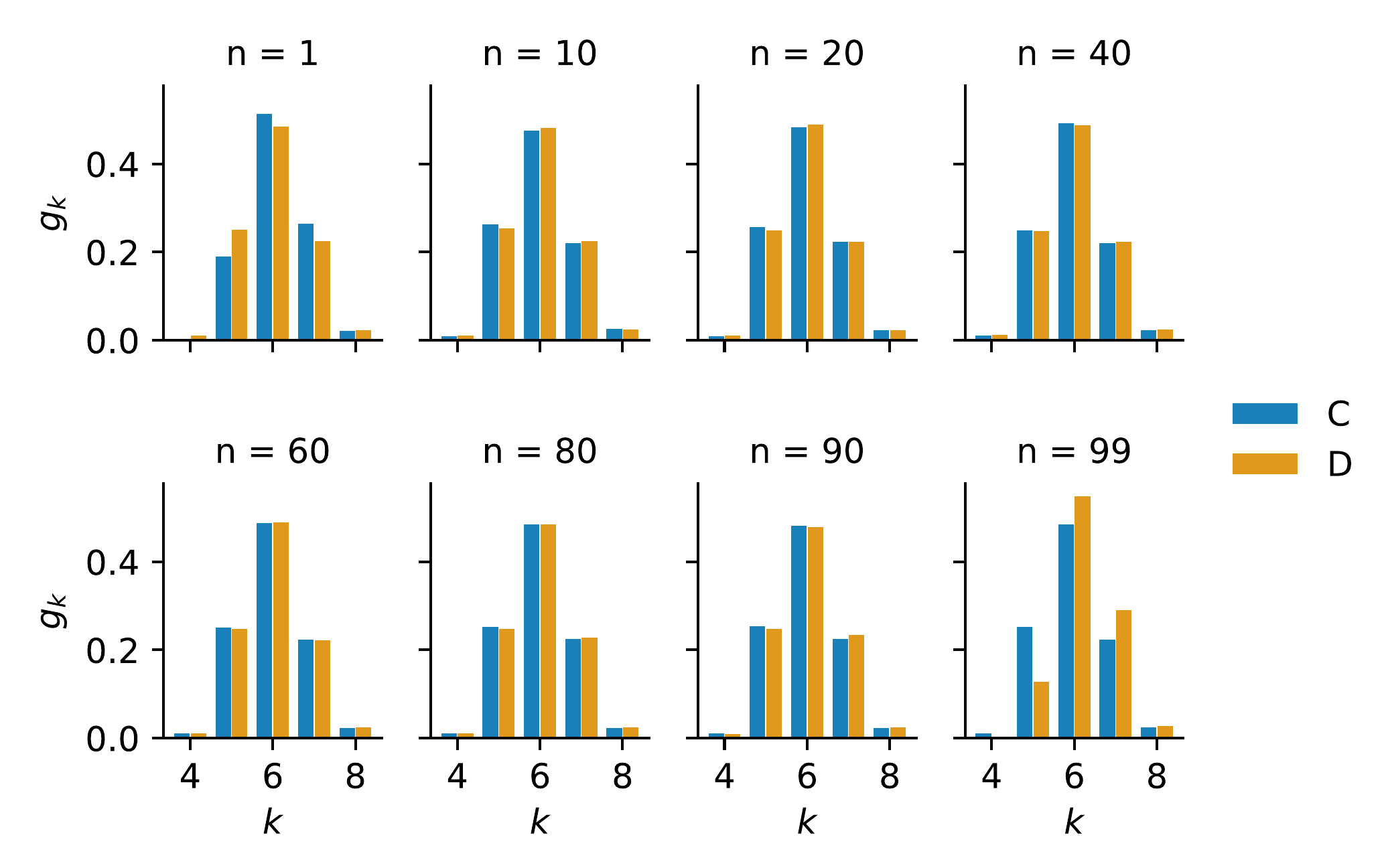}
    \caption{Neighbour distributions in the VT model for cooperators (C) and defectors (D), for varying cooperator population size, $n$. Data is generated from simulations with total population size $Z=100$ in the neutral selection limit, $\delta=0$.}
    \label{fig:compare_neighbour_dist}
\end{figure}

\Cref{fig:compare_neighbour_dist} plots neighbour distributions from simulations of the VT model for cooperators and defectors at different values of $n$. It is clear from the plot that the assumption is a reasonable one. The neighbour distributions are approximately equal for different values of $n$ and for the two cell types. The exception is when there are either very few cooperators or very few defectors, i.e.\ near $n=1$ and $n=99$ respectively. In the case where there is only one or very few cooperators, the cooperator neighbour distribution becomes slightly more narrow. The converse is true when there are few defectors.

\section{Minimising the critical benefit-to-cost ratio at which cooperation is favoured}
\label{sec:minimising_btoc_appendix}

In \Cref{sec:beneficial_coop} we considered conditions for cooperative success for a sigmoid benefit function, as defined by \Cref{eq:sigmoid_benefit}. It is clear from \Cref{fig:compare_btoc1} that the critical benefit-to-cost ratios, $(b/c)^*_1$, at which $\rho_C=\rho_D$, are minimised at $h=0.5$, and symmetric across that point. This appears to hold for both the Voronoi tessellation model with decoupled update, and for the death-birth update on a fixed hexagonal lattice. In the following we show that this is indeed true for any system where $0<s<\infty$ and the structure coefficients, $\sigma_j$, are increasing for $0\le j<k$. 

We rewrite \Cref{eq:btoc1}, defining $(c/b)^*_1$, such that cooperation is favoured for $c/b<(c/b)^*_1$
\begin{equation}
    \left(\frac{c}{b}\right)^*_1 = \frac{1}{Z-1}\sum_{j=0}^k\sigma_j\left[\beta\left(\frac{j+1}{k+1}\right)-\beta\left(\frac{k-j}{k+1}\right)\right].
\end{equation}
We have assumed that the number of neighbours, $k$, is fixed, however the results are easily generalisable to variable $k$. Defining
\begin{equation}
    \gamma_j = \beta\left(\frac{j+1}{k+1}\right)-\beta\left(\frac{k-j}{k+1}\right) \label{eq:gamma_def}
\end{equation}
we obtain 
\begin{equation}
    \left(\frac{c}{b}\right)^*_1 = \frac{1}{Z-1}\left[\sigma_k + \sum_{k>j\ge k/2} (\sigma_j-\sigma_{k-j-1})\gamma_j\right].
    \label{eq:ctob1}
\end{equation}
By taking derivatives with respect to $h$ we show that for $k/2 \ge j <k$, $\gamma_j(h)$ is maximised when $h=0.5$. In order that this corresponds to a unique maximum of $(c/b)^*_1$, and thus a minimum of the critical benefit-to-cost ratio, certain conditions on $\sigma_j$ must be satisfied.

First we show that $\gamma_j$ has one extremum at $h=0.5$ for $0<s<\infty$. We substitute \Cref{eq:sigmoid_benefit} into \Cref{eq:gamma_def} and take the first derivative with respect to $h$, letting $r=\frac{j+1}{k+1}$. Thus we obtain
\begin{align}
    \frac{d\gamma_j}{dh} &= \frac{d}{dh} \left[\frac{(1+e^{s(h-r)})^{-1} - (1+e^{s(h+r-1)})^{-1}}{(1+e^{s(h-1)})^{-1}-(1+e^{sh})^{-1}}\right] \\
    &= \frac{d}{dh} \left[\frac{e^{s(r-1)}-e^{-sr}}{1-e^{-s}} \cdot \frac{(1+e^{s(h-1)})(1+e^{sh})}{(1+e^{s(h-r)})(1+e^{s(h+r-1)})}\right] \\
    &= s\cdot\frac{e^{s(r-1)}-e^{-sr}}{1-e^{-s}} \cdot \frac{e^{sh}(1+e^{-s}-e^{-sr}-e^{s(r-1)})(1-e^{s(2h-1)})} {(1+e^{s(h+r-1)})^2(1+e^{s(h-r)})^2} .
\end{align}
Setting $d\gamma_j/dh=0$, gives one root at $h=0.5$, for $0<s<\infty$. This is a unique stationary point of $(c/b)^*_1$ so long as there is at least one value of $j\in[k/2,k)$ for which $(\sigma_j-\sigma_{k-j-1})\ne 0$. We can show that this is a maximum by considering the second derivative at $h=0.5$
\begin{align}
    \left .\frac{d^2\gamma_j}{dh^2}\right |_{h=\frac{1}{2}} = -2s^2 \cdot \frac{e^{s/2}(1+e^{-s}-e^{-sr}-e^{s(r-1)})(e^{s(r-1)}-e^{-sr})}{(1-e^{-s})(1+e^{s(r-1/2)})^2(1+e^{-s(r-1/2)})^2} 
\end{align}
which is negative given that $1/2 < r < 1$. This corresponds to $(k-1)/2<j<k$, encompassing all the values of $j$ which we sum over in \Cref{eq:ctob1}. Therefore, in order that $(c/b)^*_1$ is maximised when $h=0.5$, we require that $(\sigma_j-\sigma_{k-j-1})\ge 0$ for $k/2 \le j <k$  and non-zero for at least one value of $j$ in the range. This condition is guaranteed if $\sigma_j$ is an increasing, but not constant, function for $0\le j<k$.

\begin{figure}[!h]
    \centering
    \includegraphics[scale=1]{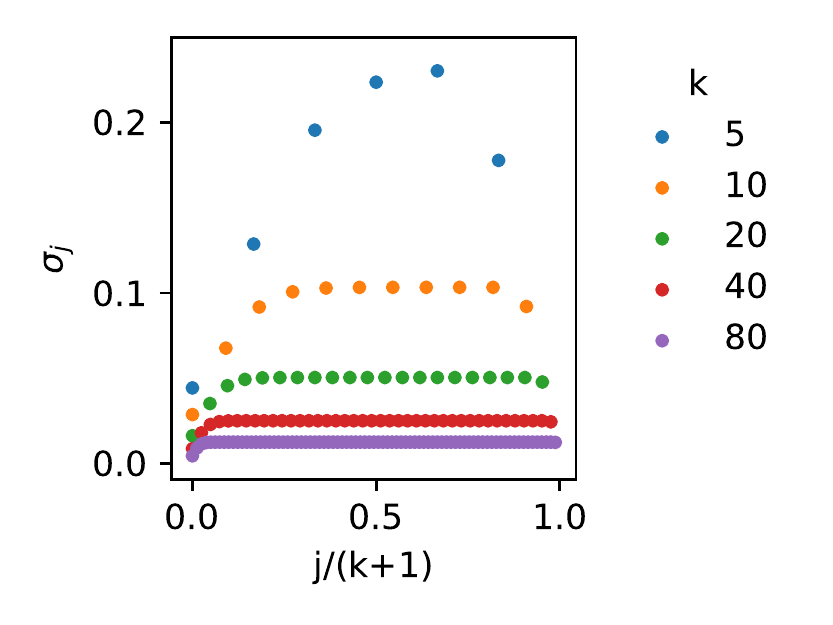}
    \caption{Structure coefficients, $\sigma_j$, for $k$-regular graphs with death-birth update rule \cite{Pena2016}. It is clear that $\sigma_j$ is increasing (or constant) for $j<k$. }
    \label{fig:sigmas_kreg}
\end{figure}

It is clear from \Cref{fig:sigma} that $\sigma_{j+1,k}>\sigma_{j,k} \;\forall j,k$ for the VT model with decoupled update, therefore $h=0.5$ maximises $(c/b)^*_1$ in this case. For $k$-regular graphs with death-birth update rule, we can verify whether this is true by using the approximate expressions for the structure coefficients derived in \cite{Pena2016}. These are plotted for various $k$ values in \Cref{fig:sigmas_kreg}. For smaller values of $k$, we can see that $\sigma_j$ is strictly increasing for $0\le j<k$. However, as $k$ increases, a growing region appears for which $\sigma_j$ is constant. So long as there is at least one value of $j<k$ for which $(\sigma_j-\sigma_{k-j-1})>0$, $(c/b)^*_1$ is maximised at $h=0.5$. However, as $k\to\infty$, we approach the case where $\sigma_j$ are constant for $j<k$, and we regain the well-mixed population result that $(c/b)^*_1$ are independent of $h$.

Thus far we have limited ourselves to the case where $0<s<\infty$. In the limit $s\to0$, we obtain an NPD game with a linear benefit function which is independent of $h$. The value of $(c/b)^*_1$ therefore does not depend on $h$ either, as can be seen in \Cref{fig:compare_btoc1}. In the limit $s\to \infty$, the VD game is approached and the benefit function ceases to be continuous. In this case the unique maximum at $h=0.5$ is maintained only if $\sigma_j$ are strictly increasing, and therefore $(\sigma_j-\sigma_{k-j-1})>0$. This is true for the VT model with decoupled update and for $k$-regular graphs with death-birth update, if $k$ is sufficiently small. On the other hand, if $(\sigma_j-\sigma_{k-j-1})=0$ for some values of $j\in[k/2,k)$, $h=0.5$ ceases to be an isolated maximum, and there is a region of $h$ values, around $h=0.5$, which maximise $(c/b)^*_1$. 

\section{Equivalence of beneficial and favoured mutants}

\label{sec:equivalence_beneficial_favoured}
\subsection{Antisymmetry-of-invasion property}
In \Cref{sec:global,sec:beneficial_coop} we derived the conditions under which a mutant is beneficial or favoured, respectively, for a global update rule. Here, we show that these conditions are equivalent if the payoffs satisfy a property we call \textit{antisymmetry-of-invasion}. We consider multiplayer games with fixed group size. However, the results can be generalised to variable group size, given certain conditions.

The values $\theta_j^A$ and $\theta_j^B$, defined by \Cref{eq:theta_fixed_k}, can be written as
\begin{equation}
    \begin{aligned}
       \theta_j^A &= \sum_{n=1}^{Z-1}(Z-n)f_j^A(n) \\
       \theta_j^B &= \sum_{n=1}^{Z-1}nf_{k-j}^A(n) \, .
    \end{aligned}
\end{equation}
Thus we have
\begin{equation}
    \theta_j^A + \theta_{k-j}^B = Z\sum_{n=1}^{Z-1}f_j^A(n) = Z\sigma_j \, ,
\end{equation}
where the last equality is from the definition of $\sigma_j$ as stated by \Cref{eq:global_update_sigma}. The condition for A to be beneficial, given by \Cref{eq:beneficial_cooperation_fixed_k}, thus becomes
\begin{equation}
    \sum_{j=0}^k \left[\theta_j^A a_j - \left(Z\sigma_j - \theta_j^A\right) b_{k-j}\right] > 0 \, .
\end{equation}
This can be rewritten in the form
\begin{equation}
    \sum_{j=0}^k\left(\theta_j^A - \frac{Z}{2}\sigma_j\right)\left(a_j+b_{k-j}\right)
        +\frac{Z}{2}\sum_{j=0}^k\sigma_j \left(a_j -b_{k-j} \right) >0 \, .
\label{eq:condition_beneficial_A_two_parts}
\end{equation}
If the payoffs satisfy
\begin{equation}
    a_j+b_{k-j}=Q \, ,
    \label{eq:symmetry_of_invasion}
\end{equation}
where $Q$ is a constant that is independent of $j$, then the first term in \Cref{eq:condition_beneficial_A_two_parts} vanishes. The condition for $A$ to be beneficial, therefore, becomes
\begin{equation}
    \sum_{j=0}^k \sigma_j \left(a_j-b_{k-j}\right)>0 \, ,
\end{equation}
which is equivalent to the condition for $A$ to be favoured, as defined by \Cref{eq:sigma_rule}. Thus the conditions for cooperation to be beneficial and favoured are equivalent when \Cref{eq:symmetry_of_invasion} holds, which we call the antisymmetry-of-invasion property. If $Q$ is independent of $k$, this result generalises to variable group size.

\subsection{Implications for antisymmetry-of-invasion}

In games which satisfy antisymmetry-of-invasion, defined by \Cref{eq:symmetry_of_invasion}, there is a fixed total payoff which can be obtained when equal numbers of $A$ and $B$ co-players are distributed between an $A$ and $B$ player. By this we mean that the $A$-player has $j$ other $A$-players in its group and $k-j$ $B$-players, whilst the $B$-player has $j$ other $B$-players, and $k-j$ $A$-players. Regardless of how the co-players are distributed (the value of $j$), the sum of the payoffs to the $A$ and $B$ player are the same. 

The implications for this property can be better understood if we consider symmetric invasion processes. Consider, for example an arbitrary evolutionary path through the state space. This path can be represented by a sequence of states
\begin{equation}
    S = (G_0,\vec{s}_0)\to (G_1,\vec{s}_1) \to \dots \to (G_L,\vec{s}_L) \, , \label{eq:path}
\end{equation}
where $G_q$ are graphs representing the population structure at time $t_q$ and $\vec{s}_q$ are $Z$-dimensional vectors giving the type of each individual at time $t_q$. Thus, $[\vec{s}_q]_i=1$ if the $i$th individual is an $A$-player and $[\vec{s}_q]_i=0$ if it is a $B$-player. Recall $Z$ is the population size. There are $L$ transitions between states, each of which is caused by an update event (i.e.\ a death and a division).

\begin{figure}[htb]
    \centering
    \begin{subfigure}{0.45\textwidth}
        \centering
        \includegraphics{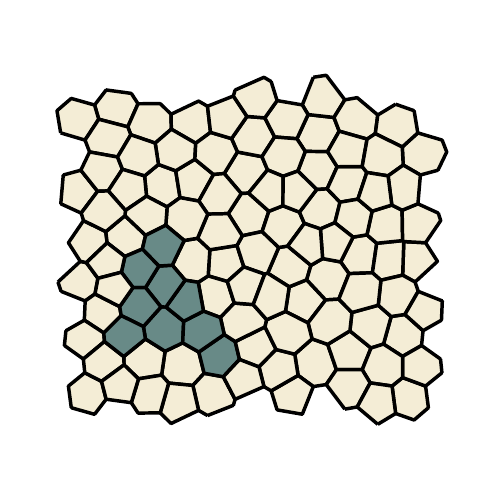}
        \caption{Original state.}
    \end{subfigure}
    \begin{subfigure}{0.45\textwidth}
        \centering
        \includegraphics{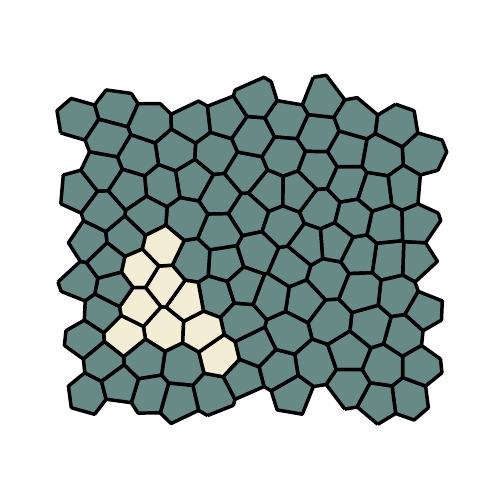}
        \caption{Symmetric state.}
    \end{subfigure}
    \caption{Symmetric states. (a) a mutant clone of $A$-players is invading a population of $B$-players. (b) a mutant clone of $B$-players is invading a population of $A$-players. If the antisymmetry-of-invasion property holds a given $A$-player in state (a) has payoff $a_j$, the equivalent $B$-player in state (b) will have payoff $b_{k-j}=Q-a_j$.}
    \label{fig:symmetric_invasion}
\end{figure}

The symmetric invasion process $\tilde{S}$ is obtained by flipping the type of each individual ($A\to B$ and $B\to A$), as illustrated in \Cref{fig:symmetric_invasion}. Thus
\begin{equation}
    \tilde{S} = (G_0,\tilde{\vec{s}}_0)\to (G_1,\tilde{\vec{s}}_1) \to \dots \to (G_L,\tilde{\vec{s}}_L) \, , \label{eq:symmetric_path}
\end{equation}
where $[\tilde{\vec{s}}_q]_i=1-[\vec{s}_q]_i$.

Given any evolutionary path $S$ and a symmetric path $\tilde{S}$ we can show that, if the antisymmetry-of-invasion property holds, the probabilities of each occurring are related in the following way:
\begin{equation}
    P(S) - P(S_0) = P(S_0) - P(\tilde{S})\, , \label{eq:path_symmetry}
\end{equation}
at least to $\mathcal{O}(\delta)$. Here, $S_0$ is the evolutionary path with neutral selection $\delta=0$, i.e.\ all individuals have the same fitness. Thus, if any given path has an advantage over the neutral process, the symmetric path must have an equivalent disadvantage. 

We can further show that the following relation between the fixation probability for an $A$-player and the fixation probability of a $B$-player, denoted by $\rho_A$ and $\rho_B$, respectively, must hold:
\begin{equation}
    \rho_A - 1/Z = 1/Z - \rho_B \, , \label{eq:fixprob_symmetry}
\end{equation}
again to $\mathcal{O}(\delta)$. Recall that $\rho_0=1/Z$ is the fixation probability for a neutral mutant. Thus, antisymmetry-of-invasion ensures that $\rho_A>\rho_0$ implies $\rho_B<\rho_0$, and hence that the conditions for $A$ or $B$ to be favourable are the same as to be beneficial.

\paragraph{Proof of \Cref{eq:path_symmetry}.} Consider a path $S$ as described by \Cref{eq:path}. The transition probability from state $(G_q,\vec{s}_q)$ to $(G_{q+1},\vec{s}_{q+1})$ is given by
\begin{equation}
\begin{aligned}
    P((G_q,\vec{s}_q)\to(G_{q+1},\vec{s}_{q+1})) 
    &= P(\vec{s}_q\to \vec{s}_{q+1}) \cdot P(G_q\to G_{q+1} | \vec{s}_q\to \vec{s}_{q+1}) \\
    &= \frac{1}{Z^2}\left\{1+\delta\left[\pi_\text{birth}(G_q,\vec{s}_q)-\pi(G_q,\vec{s}_q)\right]\right\} \cdot \psi_q
    \, , 
    \end{aligned}
\end{equation}
where $\pi_\text{birth}$ is the payoff of the proliferating individual and $\pi$ is the average payoff in the population. The probabilities for transitions between graphs are given by $P(G_q\to G_{q+1} | \vec{s}_q\to \vec{s}_{q+1}) =\psi_q$. 

The probability of $S$ occurring, given initial state $(G_0,\vec{s}_0)$, is given by multiplying the transition probabilities, i.e.\ 
\begin{equation}
    P(S) = \prod_{q=0}^{L-1}P((G_q,\vec{s}_q)\to(G_{q+1},\vec{s}_{q+1})) \, ,
\end{equation}
which in the weak selection limit $\delta\to 0$ becomes
\begin{equation}
    P(S) = \frac{1}{Z^{2L}} (1+\delta X(S))\Psi(S) + \mathcal{O}(\delta^2) \, . \label{eq:prob_path}
\end{equation}
Here,
\begin{equation}
    X(S) = \sum_{q=0}^{L-1} \left(\pi_\text{birth}(G_q,\vec{s}_q)-\pi(G_q,\vec{s}_q)\right) \, 
\end{equation}
and 
\begin{equation}
    \Psi(S) = \prod_{q=0}^{L-1} \psi_q \, .
\end{equation}

The symmetric evolutionary path $\tilde{S}$ is equivalent to $S$, except that every individual has flipped its type. We assume that, in the weak selection limit at least, graph transitions do not depend on type, and thus, $\Psi(\tilde{S})=\Psi(S)$. The payoffs of course do depend on type, thus we write
\begin{equation}
    X(\tilde{S}) = \sum_{q=0}^{L-1} \left(\pi_\text{birth}(G_q,\tilde{\vec{s}}_q)-\pi(G_q,\tilde{\vec{s}}_q)\right) \, .
\end{equation}
If the antisymmetry-of-invasion property, defined by \Cref{eq:symmetry_of_invasion}, holds then 
\begin{equation}
\begin{aligned}
        X(\tilde{S}) &= \sum_{q=0}^{L-1} \left((Q-\pi_\text{birth}(G_q,\vec{s}_q))-(Q-\pi(G_q,\vec{s}_q))\right)
        & = -X(S) \, .
\end{aligned}
\end{equation}
Therefore, substituting into \Cref{eq:prob_path}, we obtain
\begin{equation}
    P(\tilde{S}) = \frac{1}{Z^{2L}} (1-\delta X(S))\Psi(S) + \mathcal{O}(\delta^2) \, . \label{eq:sym_prob_path}
\end{equation}
Setting $\delta=0$ gives $P(S_0)=\Psi(S)/Z^{2L}$. Therefore, by summing \Cref{eq:prob_path,eq:sym_prob_path}, we obtain $P(S)+P(\tilde{S})=2P(S_0)$, from which \Cref{eq:path_symmetry} follows.

\paragraph{Proof of \Cref{eq:fixprob_symmetry}.} The fixation probability for a single initial $A$-player is obtained by summing $P(S_i)$ over all paths $S_i$ that start with a single initial $A$-player, and end with fixation for $A$-players. Summing over \Cref{eq:prob_path}, we obtain
\begin{equation}
\begin{aligned}
    \rho_A &= \sum_i \frac{\Psi(S_i)}{Z^{2L(S_i)}} + \sum_i \frac{\delta\Psi(S_i)}{Z^{2L(S_i)}}X(S_i) +\mathcal{O}(\delta^2) \\
    &= \frac{1}{Z} + \sum_i \frac{\delta\Psi(S_i)}{Z^{2L(S_i)}}X(S_i) +\mathcal{O}(\delta^2) 
    \, , \label{eq:rho_A_fix}
\end{aligned}
\end{equation}
where we have used the fact that the fixation probability for neutral selection ($\delta=0$) is $\rho_0=1/Z$. The fixation probability for $B$-players can similarly be obtained by summing $P(\tilde{S_i})$ over all paths $\tilde{S_i}$ that start from a single $B$-player and end with $B$-player fixation. Thus,
\begin{equation}
    \rho_B = \frac{1}{Z} - \sum_i \frac{\delta\Psi(S_i)}{Z^{2L(S_i)}}X(S_i) +\mathcal{O}(\delta^2) \, . \label{eq:rho_B_fix}
\end{equation}
Summing \Cref{eq:rho_A_fix,eq:rho_B_fix} gives us $\rho_A+\rho_B=2/Z$, and thus \Cref{eq:fixprob_symmetry}.




\subsection{Antisymmetry-of-invasion for public goods games in epithelia}

In \Cref{sec:epithelium} we considered the conditions under which cooperative mutants are beneficial and successful for sigmoid public goods games in the VT model with global updating and for the well-mixed population. Cooperation is beneficial when $b/c > (b/c)^*_0$ and favoured when $b/c>(b/c)^*_1$. It is is evident from \Cref{fig:coop_success}, that in general $(b/c)^*_0 \ne (b/c)^*_1$. 
However, it appears in the figure that they are equal when $h=0.5$ and/or $s\to\infty$. Recall, that $h$ is the inflection point and $s$ is the steepness of the logistic function, defined by \Cref{eq:logistic}. Here, we show that both cases satisfy the antisymmetry-of-invasion property defined by \Cref{eq:symmetry_of_invasion} and thus $(b/c)^*_0=(b/c)^*_1$ must hold.

When $s\to0$ we approach the NPD, which has a linear benefit function, given by \Cref{eq:NPD}. The cooperator and defector payoffs are thus 
\begin{equation}
       a_{j,k} = b\cdot\left(\frac{j+1}{k+1}\right) - c \qquad \text{and} \qquad b_{j,k} = b\cdot\left(\frac{j}{k+1}\right) \, ,
\end{equation}
respectively. We therefore obtain
\begin{equation}
    a_{j,k} +b_{k-j,k} = b-c \, .
    \label{eq:egfs_pgg}
\end{equation}
As $b-c$ is a constant independent of $j$ and $k$, this satisfies antisymmetry-of-invasion, defined by \cref{eq:symmetry_of_invasion}. The critical benefit-to-cost ratio above which cooperation is favoured must, therefore, be equal to the critical benefit-to-cost ratio above which cooperation is beneficial, i.e.\ $(b/c)^*_0=(b/c)^*_1$.

We can also show that antisymmetry-of-invasion is satisfied when $h=0.5$. The sigmoid benefit function, defined by \Cref{eq:sigmoid_benefit}, has the symmetry property $\beta(x)=1-\beta(1-x)$ when $h=0.5$. The cooperator and defector payoffs are therefore given by
\begin{equation}
\begin{aligned}
       a_{j,k} &= b\cdot\beta\left(\frac{j+1}{k+1}\right) - c = b\cdot\left[1-\beta\left(\frac{k-j}{k+1}\right)\right] - c \\[0.2cm]  
       b_{j,k} &= b\cdot\beta\left(\frac{j}{k+1}\right) \, ,
\end{aligned}
\end{equation}
respectively. Once again, we find that $a_{j,k} +b_{k-j,k} = b-c$. Therefore, there is antisymmetry-of-invasion when ${h=0.5}$, so $(b/c)^*_0=(b/c)^*_1$ must hold.

\end{appendices}
\end{document}